\documentclass[aps,
longbibliography,notitlepage,reprint,superscriptaddress]{revtex4-2}
\usepackage[colorlinks,linkcolor=blue,citecolor=blue,urlcolor=red]{hyperref}
\usepackage{soul}
\usepackage{changes}
\usepackage{float}
\usepackage{CJK}
\usepackage{graphicx}
\usepackage{amsmath,amssymb,amsfonts,physics}
\usepackage{siunitx}
\usepackage{soul}
\usepackage{chngcntr}
\usepackage{chemformula}
\usepackage{MnSymbol}
\usepackage{dsfont}
\usepackage{appendix}

\newcommand{\HU}{John A. Paulson School of Engineering and Applied Sciences, Harvard University, Cambridge, MA 02138, USA}
\newcommand{\HUP}{Department of Physics, Harvard University, Cambridge, MA 02138, USA}
\newcommand{\TBD}{Department of Electrical Engineering and Computer Science, Syracuse University, Syracuse, NY 13210, USA}

\begin{document}

\author{Guanhao Huang}
\email{guanhao\_huang@seas.harvard.edu}
\affiliation{\HU}
\author{Chang Jin}
\affiliation{\HU}
\author{Sophie Weiyi Ding}
\affiliation{\HU}
\author{Chaoshen Zhang}
\affiliation{\HU}
\author{Aaron M. Day}
\affiliation{\HU}
\author{Tobias Elbs}
\affiliation{\HU}
\author{Neil Sinclair}
\affiliation{\HU}
\author{Sukhad Dnyanesh Joshi}
\affiliation{\TBD}
\author{Rodrick Kuate Defo}
\affiliation{\TBD}
\author{Bertrand I. Halperin}
\affiliation{\HUP}
\author{Evelyn Hu}
\affiliation{\HU}
\author{Marko Lončar}
\email{loncar@seas.harvard.edu}
\affiliation{\HU}

\title{Ultracoherent self-assembled diamond nanomechanics reveals superfluid dynamics}

\begin{abstract}
   From gravitational-wave detection~\cite{Cole13,Cole23}, protein force microscopy~\cite{Rugar2004}, to exploration of quantum-classical boundaries~\cite{Bassi13}, many anticipated discoveries in fundamental science require improving measurement sensitivity limits. Through the fluctuation–dissipation theorem~\cite{Kubo_1966}, mechanical dissipation sets the acoustic noise for this limit. Yet, even in high-purity crystals, the microscopic mechanisms responsible for the acoustic loss remain poorly understood. Tension-induced dissipation dilution~\cite{Engelsen2024} offers a route to ultralow acoustic loss, but is challenging to implement~\cite{Field_2012} in crystalline materials~\cite{Hochreiter25,Shandilya2019,Machielse19} including single-crystal diamond~\cite{Tao2014}. Here we realize a strain-engineered diamond nanomechanical platform using a liquid-assisted van der Waals (vdW) self-assembly process~\cite{Babar2023} that harnesses intrinsic surface forces~\cite{Bhushan03} to apply tensile stress exceeding \SI{1}{GPa}. At cryogenic temperatures these resonators achieve quality factors beyond \SI{e10}{} (intrinsic material quality factors beyond \SI{e8}{}). 
   This exceptional coherence turns them into a sensitive probe for residual dissipation, elucidating three distinct two-level-system channels and one topological dissipation channel from a surface superfluid helium film~\cite{Ambegaokar80}. Our work shows how advancing mechanical coherence opens access to new regimes of physics in hybrid quantum systems~\cite{Fung24,Graham25}, precision metrology~\cite{Marletto17,Bertone2018}, and condensed-matter physics.

\end{abstract}

\maketitle

\begin{figure*}[t]
   \includegraphics[width = 1\textwidth, page = 1]{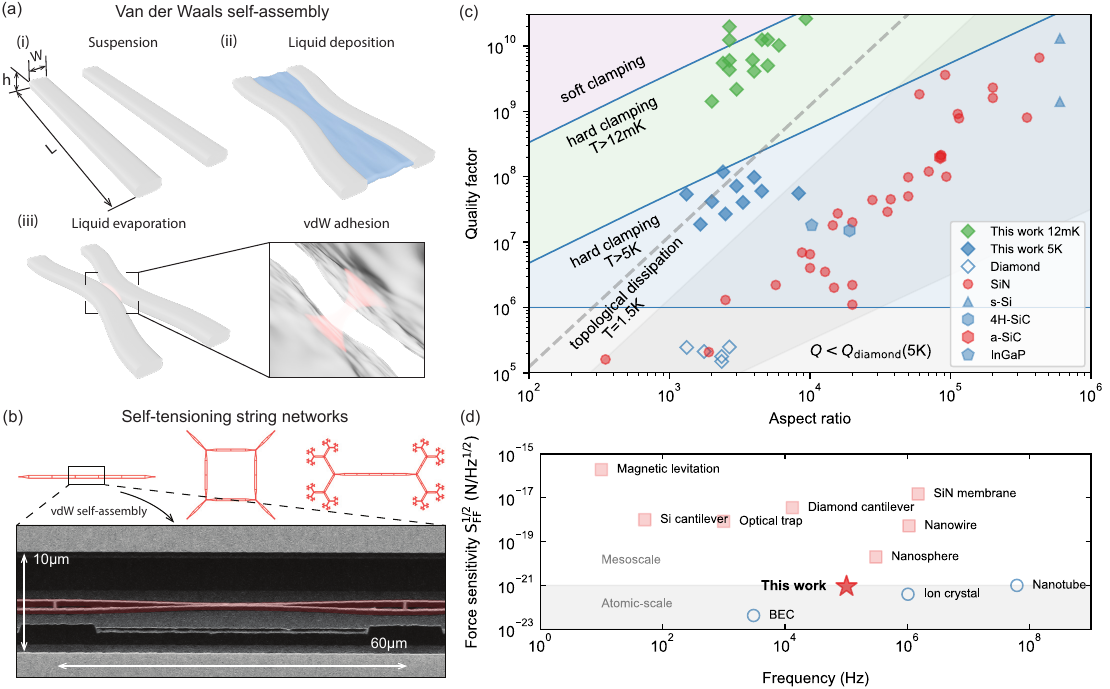} 
   \caption{\textbf{Ultralow-loss diamond nanomechanics using van der Waals (vdW) self-assembly.} \textbf{(a)} Schematic of the vdW self-assembly process: initially (i) suspended, parallel diamond nanobeams are drawn together by (ii) surface tension from liquid condensation. Upon evaporation, (iii) structural adhesion is achieved through vdW surface interactions. \textbf{(b)} False-color SEM image of segmented self-tensioning tethers enabling vdW self-assembly of complex string networks (illustrated above the SEM image) capable of supporting soft-clamped modes (two axes scaled differently). \textbf{(c)} Acoustic quality factor as a function of aspect ratio ($r=L/h$) for various nanomechanical platforms (\SI{<200}{nm} thick) down to cryogenic temperature. Amorphous (red) and crystalline (blue) platforms are shown. Shaded regions indicate typical scaling behaviors from $\propto r$ (hard clamping) to $\propto r^2$ (soft clamping)~\cite{Engelsen2024}. Theoretical projections (\SI{1}{GPa}) for diamond (100-nm thickness) at varying temperature highlight unprecedented performance achievable due to diamond's exceptional intrinsic quality factor (horizontal blue line). The estimated threshold for resolving topological dissipation in our experiment is shown as the dashed line. \textbf{(d)} Comparison of the projected force sensitivity with other mesoscale (red square) and atomic-scale (blue circles) platforms~\cite{Sudhir25}. }
   \label{fig:fig1}
\end{figure*}

\begin{figure*}[t]
    \includegraphics[width = 1\textwidth, page = 1]{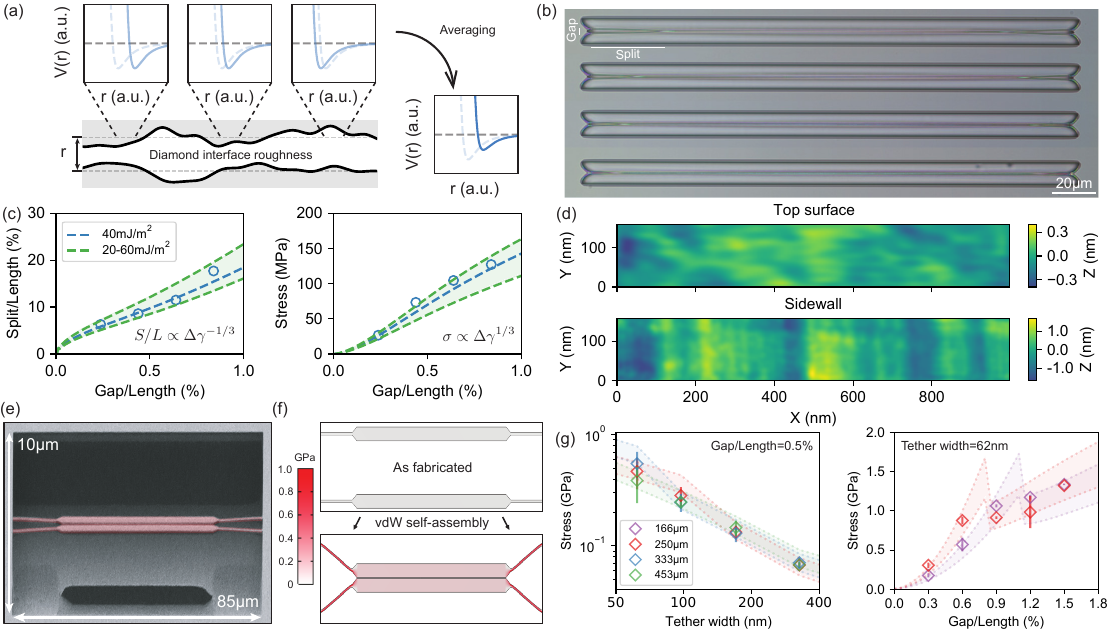}  
    \caption{\textbf{van der Waals (vdW) adhesion physics.} \textbf{(a)} Illustration showing how surface roughness reduces the effective vdW surface potential by averaging the vdW interaction across varying surface offsets. \textbf{(b)} Optical images of double-beam devices with uniform width (\SI{171}{nm}), and \textbf{(c)} experimental characterization of vdW adhesion, correlating the length of the non-adhered region (labeled ``split'' in y-axis) to the beam separation gap. Comparison with FEM simulations (dashed lines) enables estimation of the effective vdW surface potential and the corresponding sidewall roughness between \SI{0.3}{}-\SI{0.6}{nm} across devices. \textbf{(d)} Atomic force microscopy (AFM) surface topography maps of the top surface and sidewall of representative devices, with the sidewall roughness $R=\SI{0.5}{nm}$ consistent with the previous analysis. \textbf{(e)} False-color SEM image of clamp-tapered vdW self-assembled beams (two axes scaled differently), and \textbf{(f)} the finite element (FEM) schematic (not to scale). \textbf{(g)} Measured average tensile stresses of optical-grade devices in various lengths (color-coded) as a function of clamp widths and gaps, closely matching FEM predictions using vdW surface potentials between $\SI{30}{}-\SI{80}{mJ/m^2}$. Discontinuities in simulated curves correspond to decohesion events between contacting surfaces (details in Supplementary Information). }
  \label{fig:fig2}
\end{figure*}

Nanomechanical oscillators have become central tools for sensitive force microscopy~\cite{Eichler_2022} and macroscopic quantum experiments~\cite{Aspelmeyer14}, but their reach is increasingly limited by a fundamental tradeoff: the very nanoscale dimension that makes them sensitive to small signals also exacerbates acoustic losses by the surface defects~\cite{IMBODEN14}, creating a gap between bulk and nanoscale mechanical coherence. As dissipation sets the ultimate sensitivity~\cite{Kubo_1966}, this gap limits the reach of nanomechanical sensors and motivates pushing nanoscale dissipation to extreme levels—not only to improve sensor performance, but also to expose regimes of physics that are otherwise obscured by noise.

Here we show that this trade-off can be alleviated in a strain-engineered diamond nanomechanical platform. We demonstrate that our nanoscale resonators outperform bulk counterparts in both coherence and sensitivity. Our devices achieve mechanical quality factors beyond $10^{10}$ with intrinsic material quality factors beyond $10^{8}$, representing the best acoustic performance reported to date, and support projected force sensitivities at \SI{0.9}{zN/\sqrt{Hz}}, setting a new record for mesoscale systems~\cite{Sudhir25}. 
This exceptional coherence allows us to reveal the microscopic details of the loss channels with unprecedented resolution: the residual defects are dominated by distinguishable surface~\cite{Holder13} and subsurface defect complexes, rather than a featureless amorphous background. Beyond material imperfections, the same mechanical coherence makes the resonators sensitive to an entirely different class of loss, revealing a narrow dissipation feature consistent with the vortex dynamics~\cite{Ambegaokar80} of a surface superfluid film. These results illustrate how the substantial reduction of mechanical dissipation both extends the sensitivity of nanomechanical sensors and turns them into probes of microscopic and topological surface physics. Looking ahead, the projected performance directly opens access to quantum-coherent spin–mechanics interfaces~\cite{Fung24,Graham25}, nanoscale force microscopy~\cite{Rugar2004,Degen09,Eichler_2022}, and tests of quantum-classical boundary~\cite{Bassi13,Vinante20,Penrose2014}, with future up-scaling reaching the theoretical threshold for quantum tests of gravity~\cite{Marletto17}.

The performances achieved here are the result of incorporating dissipation dilution in crystalline nanomechanical resonators. By storing acoustic energy in the nearly lossless tension potential, dissipation dilution enhances the effective quality factor by orders of magnitude~\cite{Fedorov19,Engelsen2024}, allowing strained nanomechanics to surpass bulk resonators at both ambient and cryogenic temperatures. State-of-the-art implementations typically use thin-film platforms with intrinsic tensile stresses of order \SI{1}{GPa}, generated by thermal expansion~\cite{Wilson09} or epitaxial lattice mismatch~\cite{Beccari2022}. Since dissipation dilution is fundamentally material-agnostic~\cite{Engelsen2024}, efforts have turned toward crystalline materials with intrinsically lower cryogenic dissipation, including SiC, InGaP, Si, and AlN~\cite{Hochreiter25,Sementilli2025,Romero20,Manjeshwar2023,Beccari2022,Ciers24}. Unintuitively, none have yet clearly surpassed the performance of amorphous SiN resonators~\cite{Villanueva14} (see Fig.~\ref{fig:fig1}(c)), as high strain common in epitaxial layers often compromises crystal quality. Among crystalline candidates, single-crystal diamond stands out for its superior acoustic properties~\cite{Tao2014} and compatibility with color centers~\cite{CHIA2021}. However, the absence of intrinsic tensile stress in bulk diamond and its brittleness (fracture around \SI{200}{MPa}~\cite{Field_2012}) have hindered effective dissipation dilution. Although nanoscale experiments show that diamond can sustain elastic tensile stresses exceeding \SI{35}{GPa}~\cite{Amit18,Chaoqun21} without fracture, engineered tension in high-aspect-ratio devices has remained limited, as approaches based on thin films~\cite{Guo2021} or electrostatic actuation~\cite{Machielse19} introduce additional loss channels.

Here we overcome this longstanding challenge by achieving ultra-high material quality and strong dissipation dilution simultaneously in diamond nanomechanical oscillators. Our devices consist of single-crystal diamond nanomechanical strings with high-aspect-ratio that self-assemble to generate tensile strain through structural deformation (Fig.~\ref{fig:fig1}(a)). The self-assembly is achieved through a liquid-assisted~\cite{Kwok20} van der Waals (vdW) process~\cite{Babar2023} that harnesses intrinsic surface forces~\cite{Bhushan03} to impart tensile stresses up to \SI{1.3}{GPa} without added interface losses and is broadly applicable to other crystalline materials. Generalizing this approach to multi-tether networks (Fig.~\ref{fig:fig1}(b)) yields the ultracoherent devices whose performance and dissipation channels we explore in the remainder of this work. 

\section*{Van der Waals self-assembly of strained diamond nanomechanics}

Diamond is an ideal platform for ultracoherent nanomechanics thanks to its exceptionally low intrinsic acoustic damping~\cite{Tao2014} at cryogenic temperatures and chemically inert surfaces that avoid lossy native oxides. At the nanoscale, however, near-ideal crystals become sensitive to second-order effects such as residual bulk impurities, subsurface damage, and surface chemistry, so advancing their mechanical coherence requires careful control of both geometry and surfaces. We therefore begin by fabricating diamond nanostructures using quasi-isotropic undercut etching, which yields suspended beams with controlled thickness and surface properties. Using this approach, we can achieve aspect ratios beyond millimetres in length at \SI{100}{nm} thickness (process details see Supplementary Information). Because diamond already supports very high intrinsic quality factors, even moderate dissipation dilution can enable dramatic performance improvements. As shown in Fig.~\ref{fig:fig1}(c), theoretical scaling indicates that even with a compact footprint, tensile stresses on the order of \SI{1}{GPa} are sufficient to push diamond well beyond the current state-of-the-art quality factors, motivating the development of a practical method to generate large tensile strain.

To this end, we introduce a liquid-assisted van der Waals (vdW) self-assembly technique that uses intrinsic surface adhesion to strain monolithic diamond nanomechanical devices. Native to nanoscale structures and typically regarded as a nanofabrication hurdle, such forces have been widely used for strain engineering 2D materials~\cite{Dai19}. Starting from pairs of closely spaced, suspended beams, a small amount of condensed liquid pulls the beams together; upon evaporation, the beams remain adhered by vdW forces, and the resulting geometry converts the adhesion energy into tensile stress along the beam (Fig.~\ref{fig:fig1}(a)). This high-yield ($>93\%$) approach relies only on native surface forces, requires no additional post-fabrication material processing, and is therefore broadly compatible with other high-quality crystalline materials.

We quantify the effective vdW adhesion in our devices using a double-beam geometry in which the adhered length is set by the balance between adhesion and elastic restoring forces. Comparing the measured adhesion lengths to finite-element simulations (Fig.~\ref{fig:fig2}(b,c)) yields effective adhesion energies of \SI{20}{}–\SI{60}{mJ/m^2}, consistent with nanoscale sidewall roughness of \SI{0.3}{}–\SI{0.6}{nm} confirmed by atomic force microscopy (Fig.~\ref{fig:fig2}(d)). This calibration shows that, despite surface roughness (Fig.~\ref{fig:fig2}(a), details in Supplementary Information), vdW adhesion remains strong enough to generate large tensile stresses, and provides a non-destructive way to assess surface quality and adhesion energy.

Leveraging these insights, we taper the clamping regions of our beams to reduce restoring forces and concentrate strain, enabling controlled tensile stresses exceeding \SI{1.3}{GPa} as extracted from frequency measurements (Fig.~\ref{fig:fig2}(e,f,g)). The observed stress trends across beam widths and gaps are well described by a vdW-based model with adhesion energies in the range \SI{30}{}–\SI{80}{mJ/m^2}. Importantly, the adhesion interface itself introduces negligible additional loss (details see Supplementary Information), consistent with expectations from the surface-to-volume ratio of optical-grade diamond and indicating that the symmetric, high-quality adhesion interface does not degrade mechanical coherence.

\begin{figure*}[t]
    \includegraphics[width = 1\textwidth, page = 1]{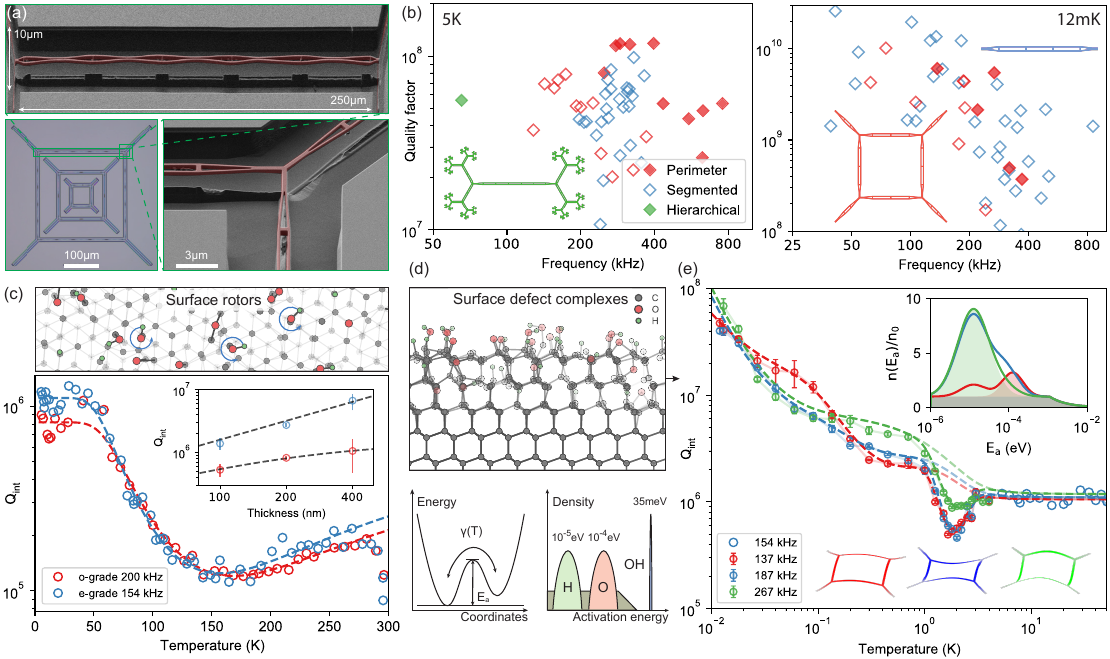}  
    \caption{\textbf{Self-assembled nanomechanical networks reveal two-level-system (TLS) defects.} \textbf{(a)} Optical and scanning electron microscopy (SEM) images of a multi-tethered diamond perimeter-mode resonator with self-tensioning capability. \textbf{(b)} Comparison of mechanical quality factors measured at \SI{5}{K} for the first few soft-clamped (solid) and hard-clamped (open) modes in perimeter-mode resonators (red), alongside the fundamental modes of hierarchical resonators (solid green) and segmented clamp-tapered resonators (open blue). With the addition of some more devices, quality factors are measured at \SI{12}{mK}, where soft-clamped modes exhibit excess loss due to shear-coupled hydrogen defects. \textbf{(c)} Temperature-dependent measurements down to \SI{5}{K} of two hierarchical resonators (100-nm thick) from electronic and optical grade diamonds, matching well with the relaxational loss from a single defect family with an activation energy of $E_a = \SI{35}{meV}$ and tunneling rate of $\Delta_0 =\SI{9.1}{MHz}$, identified as surface -OH rotors with a conceptual illustration stacked on top. Inset shows the averaged $Q_{\mathrm{int}}$ at \SI{5}{K}, showing surface-limited loss dynamics for electronic grade diamonds. \textbf{(d)} Conceptual illustrations of thermally activated defects in diamonds, including surface -OH rotors, hydrogen-induced and oxygen-induced subsurface defect complexes. \textbf{(e)} Temperature-dependent measurements of the first three modes of a perimeter mode resonator down to \SI{12}{mK}, showing distinct oscillatory behavior. Loss dynamics between 1-\SI{4}{K} is removed from the transparent theory lines. Collective fitting to a distributed TLS model reveals the presence of three defect types (see inset for defect densities): a shear-strain sensitive TLS from hydrogen-induced defect complexes (green shaded), a normal-strain sensitive TLS from oxygen-induced defect complexes (red shaded), and a uniform background of amorphous TLS (gray shaded), all at the subsurface level (more details in Supplementary Information).
    }
  \label{fig:fig3}
\end{figure*}

To fully exploit dissipation dilution, we extend vdW self-assembly from single beams to complex string-network geometries that support soft-clamped modes~\cite{Tsaturyan2017} with suppressed clamping losses. Using segmented self-tensioning tethers as modular building blocks (Fig.~\ref{fig:fig1}(b)), we induce tensile strain across multi-tether perimeter-mode and hierarchical resonators (Fig.~\ref{fig:fig3}(a)) while maintaining high yield in our scalable process. These vdW-strained networks fabricated in electronic-grade diamond exhibit clear soft-clamping enhancements at \SI{5}{K} when comparing to the hard-clamped modes at similar frequencies, with mechanical quality factors exceeding $10^{8}$, and reaching $2.6\times10^{10}$ at \SI{12}{mK} (Fig.~\ref{fig:fig3}(b)), surpassing the current state-of-the-art resonators of comparable dimensions by orders of magnitude. Together, these results establish vdW-assisted dissipation dilution in diamond as a powerful and general route to ultracoherent nanomechanical systems, and set the stage for the microscopic defect physics and topological dissipation studies discussed below.

\begin{figure*}[t]
    \includegraphics[width = 1\textwidth, page = 1]{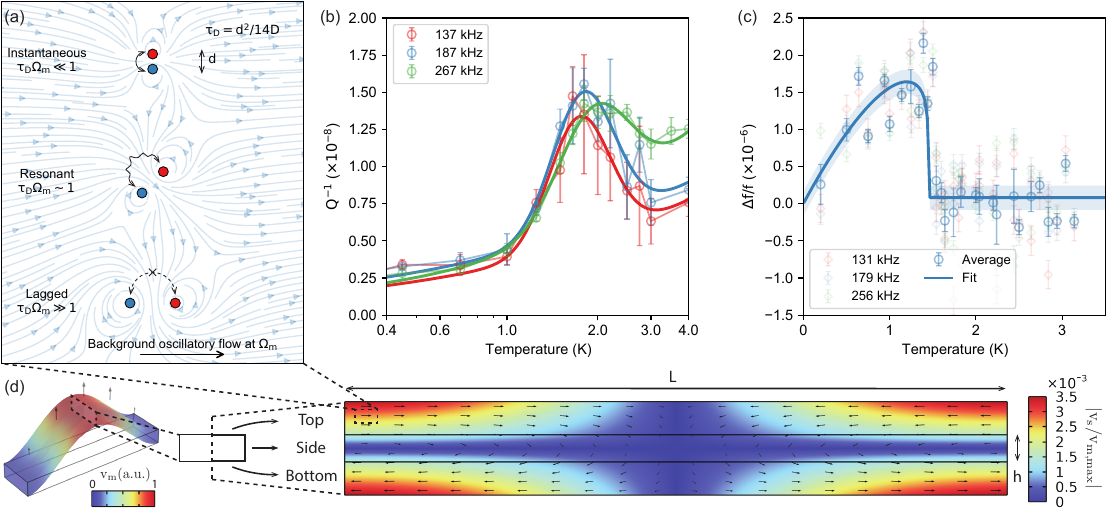}  
    \caption{\textbf{Strain-engineered diamond nanomechanics reveals superfluid physics.} \textbf{(a)} Conceptual illustration of vortex dynamics-induced damping from the \textsuperscript{4}He film on an infinite surface, where the vortex pairs switch between bound and unbound states during the topological phase transition. With an oscillating background superflow at the mechanical frequency $\Omega_m$, across the critical temperature $T_\mathrm{BKT}$, the characteristic vortex separation $d$ changes the vortex damping rate $\tau_D$, which leads to the maximized dissipation when $\Omega_m\tau_D\sim1$. \textbf{(b)} Measured excess damping induced by the topological phase transition, on the first three modes of a perimeter mode resonator, with amplitude and the shape consistent with a modified theory from the nanoscale effect of these resonator dimensions. The fitting incorporates both the vortex dynamics and a finite background contribution of TLS losses. \textbf{(c)} Measured frequency shift of the same resonator (undergone one temperature cycle) across the transition temperature at $T_c=\SI{1.42}{K}$, characteristic of the superfluid thinning by the acoustic Casimir effect of surface modes. The fit serves as a guide to the eye, capturing the far-from-transition behavior of the Casimir physics.  \textbf{(d)} FEM simulation of the out-of-plane mode of a beam structure ($L=\SI{200}{\micro\meter}$, $h=\SI{200}{nm}$, width $\SI{500}{nm}$), and the surface oscillatory superflow in the resonator frame induced by the gradient inertial force.  }
  \label{fig:fig4}
\end{figure*}

\section*{Two-level-system dynamics limits loss of diamond nanomechanics}

Two-level systems (TLS) at surfaces and interfaces are a ubiquitous source of decoherence across solid-state quantum platforms, limiting superconducting qubits~\cite{Bland2025,Odeh2025}, quantum acoustics~\cite{Wollack21,Gruenke2024,Gruenke25,MacCabe20}, and near-surface color centers~\cite{Kim15,Sangtawesin19}. Despite decades of work, most TLS remain microscopically unidentified, in part because they are ``dark'' to conventional probes: they carry no identifiable optical transition and only weak electric dipoles, so experiments typically access only an averaged response. 
By reaching a regime with exceptionally low internal loss and defect density, we enter the opposite limit where the residual damping is governed by a sparse set of elastically coupled TLS (Fig.~\ref{fig:fig3}(d)), and the resonator itself becomes a high-resolution sensor for individual defect families, with procedures described in the following paragraphs. 

All devices studied in this section use oxygen-terminated surfaces, obtained via dry or wet oxidation treatments that we find yield the highest mechanical quality factors by balancing minimal plasma damage and effective passivation (Supplementary Information). In this optimized condition, the thickness dependence of the intrinsic material quality factor (Fig.~\ref{fig:fig3}(c), inset) shows a linear scaling with surface-to-volume ratio, indicating that residual defects in electronic-grade diamond are predominantly located within a few nanometers of the surface.

The dominant temperature dependence of the material quality factor follows the characteristic form expected for thermally activated hopping between configurational minima~\cite{KRAMERS1940,Cannelli91} or quantized two-levels~\cite{Lisenfeld2015}: as temperature is swept, each defect family produces a Debye-like loss peak when its energy- and temperature-dependent relaxation rate becomes resonant with the mechanical frequency. Our measurements from \SI{10}{mK} to \SI{300}{K} thus act as a broadband “spectroscopy” of elastically coupled defects spanning many orders of magnitude in characteristic frequency (\SI{100}{MHz} to \SI{10}{THz}, details in Supplementary Information).

In our hierarchical resonators~\cite{Bereyhi2022}, both optical-grade and electronic-grade devices exhibit a pronounced, single loss peak around \SI{200}{K} (Fig.~\ref{fig:fig3}(c)) that is well described by a single defect family with an activation energy $E_a = \SI{35}{meV}$ and relaxation rate $\gamma_0/2\pi = \SI{9.1}{MHz}$. From the magnitude of the loss, we infer a surface defect density on the order of one defect per surface carbon atom (details see Supplementary Information). Together with density-functional-theory (DFT) results for hydroxylated surfaces~\cite{Holder13} and surface spectroscopy, this suggests surface –OH rotors as the origin of this high-temperature loss channel, as depicted schematically in Fig.~\ref{fig:fig3}(c,d).

At temperatures below \SI{50}{K}, the loss no longer follows a single Debye-like peak (Fig.~\ref{fig:fig3}(d,e)), indicating a more complex low-energy defect landscape. Excluding the \SIrange{1}{4}{K} region associated with superfluid helium (discussed in the next section), we perform collective fits to the temperature dependence of three mechanical modes of a single perimeter-mode resonator that exhibits different mixtures of normal and shear strain. This analysis reveals three distinct contributions: (i) a nearly uniform TLS background consistent with a thin amorphous layer at the subsurface, (ii) a defect family that couples predominantly to shear strain, and (iii) a defect family that couples predominantly to normal strain. Measurements on a control device subjected to hydrogen-plasma treatment (Supplementary Information) show a strong increase in the shear-strain sensitive channel and in the amorphous background, allowing us to attribute the shear-sensitive TLS to hydrogen-induced subsurface complexes, while the normal-strain-sensitive TLS are associated with oxygen-related subsurface defects (Fig.~\ref{fig:fig3}(d,e)). Details on the properties of these defects and the most likely candidates are presented in Supplementary Information.

Beyond the immediate gains in cryogenic coherence, this microscopic identification of surface –OH rotors and hydrogen- and oxygen-induced subsurface defect complexes establishes low-frequency nanomechanics as a general tool for surface and defect physics. Our measurements provide quantitative benchmarks--activation energies, defect densities, and strain symmetries--that can be directly compared with density-functional calculations of candidate complexes, providing a quantitative bridge between macroscopic dissipation and microscopic defect structures. The same strategy of combining extreme mechanical coherence, geometry control and temperature spectroscopy can be ported to other crystalline platforms, offering a route toward systematically resolving the TLS landscape that limits a wide range of quantum devices.
More broadly, these results illustrate how pushing mechanical dissipation to extremely low levels turns nanomechanical resonators into sensors for microscopic physics, setting the stage for the observation of entirely different loss mechanisms--such as the topological dissipation from a surface superfluid film discussed in the following section.

\section*{Evidence for topological dissipation from surface superfluid}

So far, we have focused on loss channels intrinsic to the diamond. A distinct feature in the temperature range between \SI{1}{K} and \SI{4}{K} requires a different explanation. Its origin cannot be accounted for by material defects: the ratio between the excess dissipation and the accompanying frequency shift differs by nearly two orders of magnitude from what is expected for any internal relaxational mechanism, where the two are related as the imaginary and real parts of a single susceptibility. 
Consequently, we attribute these effects to two different mechanisms. We find that the damping effects 
can be understood as arising from motion of quantized vortices in a thin superfluid helium film coating the device surfaces, in the vicinity of a topological phase transition ~\cite{Bishop78,Bishop80,Nelson77}.
On the other hand, we believe that the observed increase in the resonant frequencies below \SI{1.42}{K} (Fig.~\ref{fig:fig4}(c))   most likely reflects a thinning of the helium film arising from the Casimir effect of order-parameter fluctuations in the superfluid phase~\cite{Zandi04}, although important details of the latter observations remain to be explained.

In bulk three-dimensional (3D) \textsuperscript{4}He, the superfluid transition is a conventional continuous phase transition in which the complex order parameter develops true long-range phase coherence. In a 2D film, by contrast, continuous symmetry breaking at finite temperature is forbidden by the Hohenberg-Mermin–Wagner theorem~\cite{Hohenberg67,Mermin66,Berezinskii:1970pzv,Berezinskii:1972fet}, and the system instead undergoes a topological Berezinskii–Kosterlitz–Thouless (BKT) transition~\cite{Kosterlitz_1973}. This transition is driven by the binding and unbinding of vortex–antivortex pairs in the superfluid film (Fig.~\ref{fig:fig4}(a)), and the dynamics of these vortex pairs produce a narrow dissipation peak slightly above the critical temperature $T_\mathrm{BKT}$ as their characteristic damping time becomes comparable to the drive period~\cite{Ambegaokar80,Ambegaokar78}, absent in the conventional continuous phase transition~\cite{Yauhen19}.

Experimentally, we observe an excess loss peak of magnitude \(Q^{-1} \sim 10^{-8}\) near \SI{2}{K} in the first three modes of a perimeter-mode resonator (Fig.~\ref{fig:fig4}(b)), on top of the TLS background discussed in the previous section. The theoretical fit to the shape of this peak leads to an effective vortex core size $a\approx\SI{3}{nm}$ with diffusivity $D\approx\SI{e-8}{m^2/s}\sim\hbar/m_\mathrm{He}$ and a separation for the dominant vortex pairs of order $a e^l$, with $ l \approx 4$.   This puts one in a regime where the vortex dynamics is effectively modified by the small lateral size of the resonator~\cite{Kotsubo86}, qualitatively different from the situation in a planar geometry
~\cite{Bishop80} (the appropriately modified theory is discussed in Supplementary Information). The superfluid motion is obtained both by the finite-element analysis and an analytical calculation for a long beam with a rectangular cross-section, in which the superfluid component flows along the beam surface in response to the oscillatory inertial force, while the viscous normal component is effectively pinned to the substrate.  We find that the longitudinal gradient of the inertial force drives superflow primarily parallel to the axis of the beam, with velocities in opposite directions on the top and bottom surfaces (see Fig.~\ref{fig:fig4}(d) and details in the Supplementary Information). 
This produces a damping peak when the vortex pair density maximizes at a separation close to the transverse size of the resonator. 

Because the relevant superflow pattern becomes weaker as the device aspect ratio $r$ increases, the resulting superfluid-induced loss scales as \(Q_s^{-1} \propto r^{-2}\). Even in optimally soft-clamped resonators, dissipation dilution cannot compensate for this scaling, so observing the BKT-induced damping in flexural motion requires the extremely low intrinsic loss achieved in our vdW-strained diamond devices.

The accompanying frequency shift across the transition (Fig.~\ref{fig:fig4}(c)), on the order of a part in \SI{e6}{}, is substantially larger than what would be expected from the dynamical susceptibility that produces the loss peak~\cite{Bishop78,Bishop80}. This indicates that more subtle static effects of the superfluid film are now visible. 
The observed resonance frequency shift below 1.42K is consistent with a change in the resonator mass due to a decrease in the helium film thickness of $\sim 0.03$ monolayers. A decrease of this magnitude would be consistent with observations in thicker films, where a thickness difference of order 2\% between the superfluid and normal-fluid phases has been found, which has been attributed to a fluctuation-induced pressure (acoustic-Casimir effect) on the film~\cite{Zandi04,Ganshin06}. 
In this scenario, thermally excited surface acoustic modes in the superfluid phase (quasi-long range phase-coherent) modify the effective van der Waals potential between the helium and the diamond surface by a thickness-independent fraction proportional to $k_B T / A_H$, where $A_H$ is the Hamaker constant of the diamond–He interface~\cite{Ganshin06}. This change leads to a slight thinning of the film and hence an increase in the mechanical frequency as the system crosses into the superfluid phase. 
As the temperature is lowered further and thermal fluctuations are suppressed, the thermal Casimir contribution diminishes and the film thickness—and thus the frequency—returns toward its original value, just as we observed. 

Far from the transition, the magnitude of the inferred film-thickness change is consistent with acoustic-Casimir thinning seen in thicker helium films. However, near $T_\mathrm{BKT}$ this simplified picture breaks down: we observe an abrupt frequency drop immediately above the transition, whereas the high-frequency fluctuations responsible for Casimir thinning should evolve smoothly and persist to temperatures well above $T_\mathrm{BKT}$. A discontinuous thinning would instead resemble a first-order transition, not the expected BKT behavior; we therefore treat Casimir thinning as setting the correct scale but not the near-transition behavior, with alternatives discussed in the Supplementary Information.

The evidence for BKT-related topological dissipation and Casimir-induced film thinning in our nanomechanical resonators highlights their ability to probe phase transitions that are otherwise hidden in conventional structures. Because the superfluid contribution to the loss is small for high-aspect-ratio structures, it would be masked in devices with higher intrinsic dissipation. Here, the combination of vdW-engineered strain and ultralow material loss converts the resonator into a sensitive detector of the dynamics and fluctuations of an external quantum fluid, establishing vdW-strained diamond nanomechanics as a powerful probe of emergent physics in low-dimensional quantum matter.

\section*{Conclusions}

We have introduced a strain-engineering approach for diamond nanomechanical systems that leverages van der Waals (vdW) self-assembly to generate tensile stresses beyond \SI{1}{GPa} while preserving intrinsic crystal quality. This enables mechanical quality factors beyond $Q = 10^{10}$ with material quality factors beyond $Q_\mathrm{int} = 10^{8}$, corresponding to quantum-coherence times of \SI{18}{s} (lifetime-limited, dephasing see Supplementary Information) and projected force sensitivities $S_F^{1/2} = $\SI{0.9}{zN/\sqrt{Hz}}, enabling quantum control at free-space incident light below \SI{1}{\micro\watt} without the need for an optical cavity. These performances set a new benchmark for mesoscale platforms and place vdW-strained diamond resonators in the regime where force-based detection of individual nuclear spins within their coherence times becomes feasible in realistic field gradients~\cite{Rugar2004,Degen09,Eichler_2022}. This capability is particularly impactful for magnetic resonance force microscopy, where single-nuclear-spin sensitivity is a long-standing requirement for three-dimensional protein imaging. In this regime, nuclear spin detection must contend with quantum measurement backaction at the level of individual spins, while simultaneously opening a path toward quantum-coherent nuclear spin–mechanical interfaces~\cite{Fung24,Graham25}.

More broadly, our results show that once mechanical dissipation is substantially reduced, thus dramatically reducing defect densities, nanomechanical resonators can become a generic spectroscopy tool for ‘dark’ surface and subsurface defect complexes that lack optical~\cite{Doherty13} or electrical signatures~\cite{Lisenfeld2015}, providing information that is difficult to obtain by any other technique. 
Our findings provide long-ranging implications not only for quantum acoustic devices~\cite{Wollack21,Gruenke2024,Gruenke25,MacCabe20} across a broad frequency range, but also for the broader efforts to understand and control surface defects in quantum materials~\cite{Kim15,Sangtawesin19,Bland2025,Odeh2025}, where TLS-limited coherence remains a central challenge. 
For example, identification of the relaxational coupling to a common, strain-selective TLS would enable multi-mode noise correlation and suppression of TLS-induced decoherence in quantum devices~\cite{Lidar2003}. 
Further, we uncover and analyze a qualitatively different channel, providing experimental evidence for topological dissipation from the BKT transition of a surface superfluid helium film~\cite{Ambegaokar78}, together with a Goldstone–Casimir thinning effect~\cite{Zandi04}. These observations highlight high-aspect-ratio diamond nanomechanics as a powerful new probe of low-dimensional quantum matter.

Looking ahead, combining vdW-enabled dissipation dilution with increased resonator–undercut depth and cm-scale diamond chips should yield aspect ratios comparable to state-of-the-art SiN devices~\cite{Cupertino2024}, offering realistically another $\sim10^{5}$ enhancement in dilution factors, with quality factors reaching beyond $\SI{e15}{}$. As we discuss in the Supplementary Information, better cryogenic refrigeration and higher aspect ratio will enable performance  reaching threshold identified for quantum tests of gravity through gravitationally-induced entanglement~\cite{Marletto17}, quantum-classical boundary reaching the GRW point~\cite{Bassi13,Vinante20}, and new bounds for ultralight dark matter~\cite{Bertone2018,Manley21}, made possible by transforming vdW stiction--often a fabrication hurdle--into a powerful tool for nanoscale science.

\section*{Acknowledgments}
We thank Nils J. Engelsen, Yiqi Wang, Katie Barajas, Renee George, Michael Haas, and Yan-Qi Huan, Jiangdong Deng, Ling Xie for helpful discussions. This work was supported by AFOSR (Grant No. FA9550-23-1-0333), AWS (Grant No. A50791), and G.H. gratefully acknowledges financial support from the Swiss National Science Foundation (Postdoc.Mobility, grant number 222257), and Harvard’s Aramont Fellowship for Emerging Science Research.  A.M.D. acknowledges support from the Harvard Quantum Initiative Postdoctoral Fellowship in Science and Engineering and the AWS Center for Quantum Networking. N.S. acknowledges support from NSF Center for Quantum Networks (ERC EEC-1941583). This work was performed in part at the Harvard University Center for Nanoscale Systems (CNS); a member of the National Nanotechnology Coordinated Infrastructure Network (NNCI), which is supported by the National Science Foundation under NSF award no. ECCS-2025158.

\section*{Contributions}
G.H. conceived the idea for the project and designed the devices. G.H., C.J. analyzed the data. G.H., S.W.D. C.Z. fabricated the devices. G.H., C.J., S.W.D., C.Z., A.M.D., N.S., and T.E. contributed to the experimental setup and sample measurement. S.D.J. and R.K.D. performed the DFT simulation. G.H., C.J., B.I.H., and E.H. performed the theoretical modeling. M.L., G.H. supervised the project. All authors contributed to the discussion and writing of the manuscript. 

\appendixpage
\appendix
\setcounter{tocdepth}{2}
\counterwithin{figure}{section}

\section{Device fabrication and optimization}\label{app:fab}

Suspension methods previously explored for diamond nanostructures include focused-ion beam milling~\cite{Babinec11}, thin-film transfer methods~\cite{Babinec11,Guo2021}, angled etching~\cite{Burek2014}, and quasi-isotropic etching~\cite{Khanaliloo2015}. Among these, quasi-isotropic etching uniquely provides precise three-dimensional control over device geometry—including slab thickness through adjustments in beam and trench widths—critical for nanoscale strain engineering. 

We fabricate our devices from single-crystal diamond substrates sourced from Element Six in two grades: electronic-grade (nitrogen impurity \SI{<1}{ppb}, dislocation density $\sim$\SI{e4}{cm^{-2}}), representing the purest available diamond ideal for assessing fundamental loss limits; and more readily available optical-grade (nitrogen impurity \SI{200}{}-\SI{300}{ppb}), providing a lower bound on optimal mechanical performance. 

\begin{figure}[t]
    \includegraphics[width = 0.48\textwidth, page = 1]{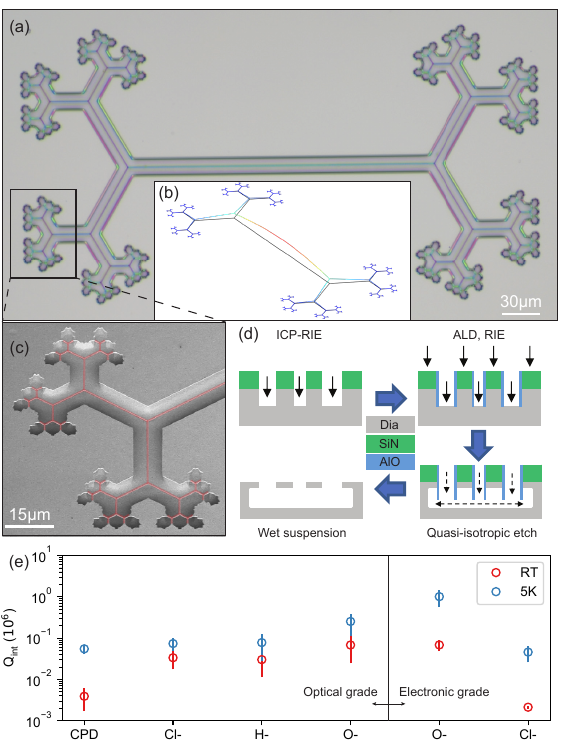}  
    \caption{\textbf{(a)} Optical image of a suspended hierarchical diamond string oscillator with up to six branchings, accompanied by \textbf{(c)} a false-color SEM image detailing the suspended nanomechanical strings within isotropically etched diamond trenches. \textbf{(b)} Finite element modeling (FEM) of the fundamental vibrational mode illustrating a soft-clamped displacement profile under zero tension, reducing radiation losses caused by clamp overhang by more than 100-fold. \textbf{(d)} Simplified fabrication flow starts with etching nanoscale features in a bulk diamond crystal using SiN hardmask. Next, atomic layer deposition (ALD) is used for sidewall protection, followed by quasi-isotropic oxygen plasma etching to undercut the structures, enabling the suspension of high-aspect-ratio slab-like nanomechanical string devices from single-crystal diamond plates. \textbf{(e)} Comparison of intrinsic quality factor $Q_{\mathrm{int}}$ for 100-nm-thick devices under different surface finishes--critical point drying, chlorine plasma, hydrogen plasma, and oxygen plasma--measured at room temperature (RT) and 5K for both optical grade and electronic grade diamond. 
    Using X-ray photoelectron spectroscopy (XPS) spectra (Fig.~\ref{fig:SI_Fig1}), we found that oxygen plasma treatment offers the optimal balance between plasma-induced damage and surface passivation, achieving the highest $Q_{\mathrm{int}}$ overall.  }
  \label{fig:SI_fabflow}
\end{figure}

The fabrication process flow is shown in Fig.~\ref{fig:SI_fabflow}, and begins by coating the diamond substrates with a 200-nm PECVD silicon nitride (SiN) layer used as a hard mask in future steps, followed by spin-coating with \SI{340}{nm} ZEP-520A resist. A thin conductive layer (5-nm gold) is deposited via electron-beam evaporation to reduce charging during electron-beam lithography. After lithographic exposure, the gold layer is chemically removed, and the ZEP resist is developed using N50 developer. The pattern is subsequently transferred into the SiN hard mask via reactive ion etching (RIE) using a $\mathrm{CF_4/C_4F_8}$ gas mixture. ZEP residues are then stripped by overnight immersion in PG remover at 80°C. Oxygen plasma vertical RIE is used to etch the pattern \SI{800}{nm} deep into the diamond substrate.

To preserve sidewall smoothness during the critical isotropic undercut step, a conformal 40-nm alumina coating is applied using atomic layer deposition (ALD). A brief argon-chlorine RIE step selectively removes the top alumina layer, exposing the diamond surface for subsequent undercutting. High-power isotropic oxygen plasma etching then undercuts and releases the diamond devices, resulting in a controlled 5-\si{\micro\meter} trench beneath each structure, which prevents unintended adhesion to the trench floor. Device thicknesses and uniformity are verified via scanning electron microscopy (SEM). The alumina protection ensures exceptionally smooth sidewalls necessary for consistent and reproducible van der Waals (vdW) adhesion.

\begin{figure}[t]
    \includegraphics[width = 0.495\textwidth, page = 1]{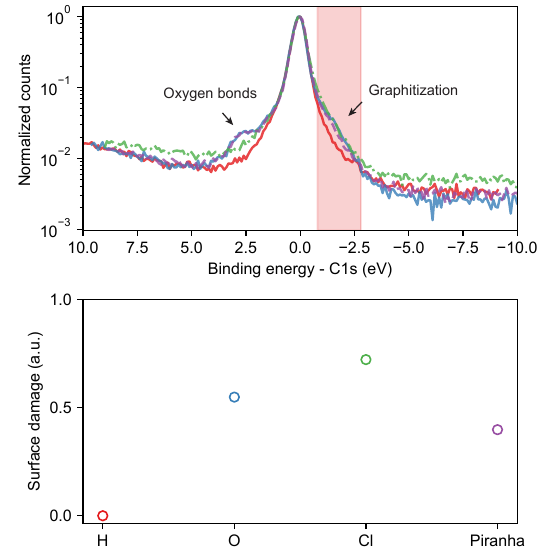}  
    \caption{X-ray photoelectron spectroscopy (XPS) spectra (color-coded) of different surface treatment (color-coded, see lower panel), centered on the carbon C1s peak, highlighting characteristic C-O bonding (\SI{2}{eV}) and plasma-induced graphitic (sp2 carbon) damage (\SI{-1}{eV}, red shaded area). The integrated area of the sp² carbon component, normalized to the hydrogen termination, is shown in the lower panel. Oxygen plasma treatment offers the optimal balance between plasma-induced damage and surface passivation, achieving the highest $Q_{\mathrm{int}}$ overall. No significant change in $Q_{\mathrm{int}}$ was observed before and after Piranha cleaning, except for partial recovery from plasma damage.}
  \label{fig:SI_Fig1}
\end{figure}

Post-undercut cleaning involves a 2-\si{min} hydrofluoric acid (HF) etch followed by a 3-\si{min} hot Piranha solution treatment to remove residual contamination. Devices then undergo a final surface-finishing step, categorized as either wet or dry. The wet finish typically employs aqueous chemical treatments followed by an IPA rinse and blow-drying, while the dry finish involves critical-point drying (CPD), followed by brief oxygen plasma cleaning to remove CPD contaminants, and subsequent plasma-induced surface termination. 

\subsection*{Surface treatment discussions}
Surface finishing methods were systematically characterized using X-ray photoelectron spectroscopy (XPS) to correlate surface chemistry with mechanical quality factors (Fig.~\ref{fig:SI_fabflow}(e), XPS data in Fig.~\ref{fig:SI_Fig1}). We observed that CPD alone introduces significant contamination, resulting in the lowest quality factors. In contrast, oxygen-based finishes (via either plasma or wet oxidation treatments such as Piranha or tri-acid) yield optimal results for both diamond grades, with wet oxidation exhibiting marginally higher performance due to minimized surface graphitization. Oxygen termination provides hydrophilic passivation, creating an adsorbed water layer that protects against further surface contamination, as confirmed by XPS showing surface hydroxyl (-OH) groups.

While XPS data indicate hydrogen plasma effectively removes surface graphitization, mK-temperature measurements indicate proliferation of hydrogen-induced defects at lower activation energies, thus leading to overall reduced mechanical quality factors. Chlorine plasma treatment induces the most pronounced lattice damage, particularly severe in electronic-grade diamond. This damage manifests as significantly reduced mechanical quality factors and pronounced Duffing softening nonlinearities, indicating lattice expansion and doping by chlorine ions. The comparatively better resilience of optical-grade diamond under chlorine treatment suggests that higher impurity levels may mitigate chlorine-related doping by occupying available vacancy sites.

We observe that when the device thickness approaches \SI{50}{nm}, non-uniformity of the beam leads to weak points at random locations, causing many beams to break and collapse onto the substrate surface. Such an effect still persists even with a gentle thinning process such as atomic layer etching~\cite{Kanarik2018}. This issue suggests that thin-film platforms with much better thickness uniformity are required to approach \SI{10}{nm} of device thickness, which is commonly utilized by the SiN platform.

\subsection*{Atomic layer etching of diamond}

To minimize surface-induced ion damping and suppress plasma-induced disorder on single-crystal diamond nanomechanical resonators, we employ an oxygen–argon atomic layer etching (ALE) sequence. Differing from conventional processes~\cite{Fischer23} that employ chlorine, we avoid chlorine as it has been shown to induce lattice expansion by surface diffusion. The ALE process consists of alternating low-energy oxygen activation steps, which functionalize and partially oxidize the topmost carbon layer, and Ar desorption pulses, which selectively remove the activated carbon in a quasi layer-by-layer fashion. When operated in the appropriate window of bias voltage and plasma power, the etch progression exhibits a self-limiting plateau characteristic of true ALE, wherein each cycle removes approximately one chemically modified monolayer and additional ion exposure produces no further material loss. Intermediate-energy conditions produce this saturation behavior, whereas higher-energy conditions transition into conventional sputtering, yielding substantially higher removal rates. As extracted from SEM-based sidewall measurements, the etch rate increases from $\sim 1.0$~\AA/cycle in the self-limiting regime to $\sim 3.5$~\AA/cycle under a 3~V bias, and further increases to $\sim 8.5$~\AA/cycle under a 14~V bias. Roughness-wise, the sputtering regime produces a smooth surface, whereas etch rates in the single-atomic-layer regime show surface roughening, characteristic of chemical etching by oxygen, which displays crystallographic facets.

The morphological evolution of the devices during ALE reflects strong crystallographic anisotropy in the etching chemistry. Transmission electron microscope (TEM) images, shown in Fig.~\ref{fig:SI_TEM}, reveal that the etch front advances preferentially along diamond lattice planes oriented approximately $30^\circ$ from the surface, originating from the sharp top edges of the suspended beams. This facet-selective retreat narrows the beam while maintaining well-defined edges. This behavior suggests that oxygen activation and subsequent Ar-mediated desorption are more efficient on these inclined crystal planes, and is consistent with the focused ion beam imaging result of the underside of the device, which had been chemically etched by the quasi-isotropic oxygen plasma.

Following the final Ar desorption step, the diamond surface is rendered carbon-terminated. Because the last step ejects oxygen-bound carbon complexes, the steady-state O coverage is significantly reduced compared with that resulting from standard oxygen plasma etches. The resulting oxygen-depleted termination is expected to suppress the formation of sp$^2$-like reconstructions and dangling-bond states, thereby reducing surface TLS channels. However, to clean the surface contamination due to sputtered Si atoms, we use HF followed by Piranha to clean the sample as a finishing step. In combination with its crystallographic selectivity and monolayer precision, this ALE approach provides a robust pathway for preparing ultra-clean diamond surfaces optimized for quantum sensing applications.

In the experiment, we observed a significantly higher material quality factor from the ALE-treated samples, shown in Table.~\ref{table:0} and Table.~\ref{table:1}. Based on the device thickness determined by the SEM and the resonant frequency of the device, we can remove the dissipation dilution effect and calibrate the intrinsic material quality factor. The best device exhibits a material quality factor around \SI{6e8}{}. However, there is substantial variation in quality factors and frequencies among ALE-treated devices, with possible contributions from acoustic radiation losses. Therefore, we want to bound the intrinsic quality factor rather conservatively, with $Q_\mathrm{int,ALE}>$\SI{e8}{}. For the dimensions of the best devices shown in Fig.1(c) of the main text, see Table~\ref{table:0}.

\begin{table}[t]
\begin{center}
\caption{Summary of the device dimensions in Fig.1(c). $\sim$ indicates thickness nonuniformity and only represents the thickness of the horizontally oriented beams. }
\label{table:0}
\begin{tabular}{ | c | c | c| c | c |  } 
  \hline
  Design & L (\SI{}{\micro\meter}) & Gap ratio (\%) & h (nm) & Q \\ 
  \hline\hline
  Double beam & 166 & 1.5 & 100 & 19e6\\
  \hline
  Double beam & 250 & 1.5 & 100 & 27e6\\
  \hline
  Double beam & 333 & 0.5 & 100 & 40e6\\
  \hline
  Double beam & 453 & 0.5 & 100 & 60e6\\
  \hline
  Perimeter & 240 & 1 & $\sim$100 & 118e6\\
  \hline
  Perimeter & 132 & 1 & $\sim$100 & 54e6\\
  \hline
  Segmented & 400 & 1.5 & $\sim$100 & 98e6\\
  \hline
  Segmented & 300 & 1.5 & $\sim$100 & 72e6\\
  \hline
  Segmented & 200 & 1.5 & $\sim$100 & 42e6\\
  \hline
  Hierarchical & 830 & 1 & $\sim$100 & 56e6\\
  \hline
  Perimeter & 240 & 1 & $\sim$100 & 5.47e9\\
  \hline
  Perimeter & 390 & 1 & $\sim$100 & 6.10e9\\
  \hline
  Segmented & 400 & 1. & $\sim$100 & 4.09e9\\
  \hline
  Segmented & 300 & 1. & $\sim$100 & 2.17e9\\
  \hline
  Segmented & 200 & 1. & $\sim$100 & 1.42e9\\
  \hline
  Segmented & 1000 & 1.5 & $\sim$200 & 12.5e9\\
  \hline
  Segmented & 1000 & 1.5 & $\sim$200 & 4.99e9\\
  \hline
  Segmented & 1000 & 1.5 & $\sim$200 & 12.4e9\\
  \hline
  Seg-ALE & 400 & 0.5 & $\sim$150 & 19.7e9\\
  \hline
  Seg-ALE & 400 & 0.5 & $\sim$150 & 5.99e9\\
  \hline
  Seg-ALE & 400 & 0.5 & $\sim$150 & 12.3e9\\
  \hline
  Seg-ALE & 400 & 0.5 & $\sim$150 & 4.32e9\\
  \hline
  Seg-ALE & 1400 & 1 & $\sim$150 & 26.1e9\\
  \hline
  Peri-ALE & 900 & 1 & $\sim$150 & 10.2e9\\
  \hline
  \hline
\end{tabular}
\end{center}
\end{table}

\section{Mechanical loss characterization}\label{app:cha}

\begin{figure*}[t]
    \includegraphics[width = 1\textwidth, page = 1]{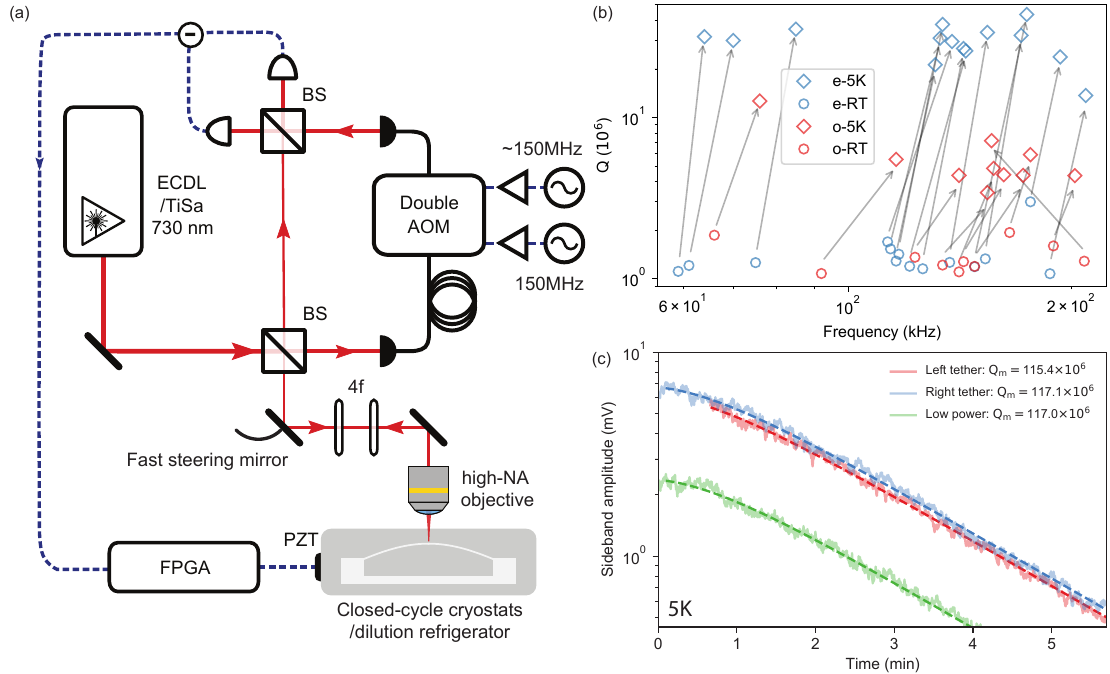}  
    \caption{\textbf{(a)} Simplified scheme of the nanomechanics characterization setup. ECDL: external cavity diode laser, BS: beam splitter, AOM: acousto-optic modulator, PZT: piezoelectric actuator, FPGA: field-programmable gate array. \textbf{(b)} Comparison of mechanical quality factors for co-fabricated electronic-grade and optical-grade devices measured at room temperature (RT) and \SI{5}{K}. All devices are 400-nm thick, under light tensile stress (\SI{25}{MPa}), and span lengths from \SI{300}{\micro\meter} to \SI{2}{mm}. The observed quality factors exhibit minimal dependence, strongly indicating that gas damping is not the dominant loss mechanism. \textbf{(c)} Mechanical quality factors measured on a soft-clamped mode of a perimeter-mode resonator as a function of laser beam position and input power. This measurement is performed in the Attodry cryostat at \SI{5}{K}. The results show consistency within 2\%, ruling out significant photothermal or optical backaction effects.}
  \label{fig:SI_Fig2}
\end{figure*}

\begin{figure*}[t]
    \includegraphics[width = 1\textwidth, page = 2]{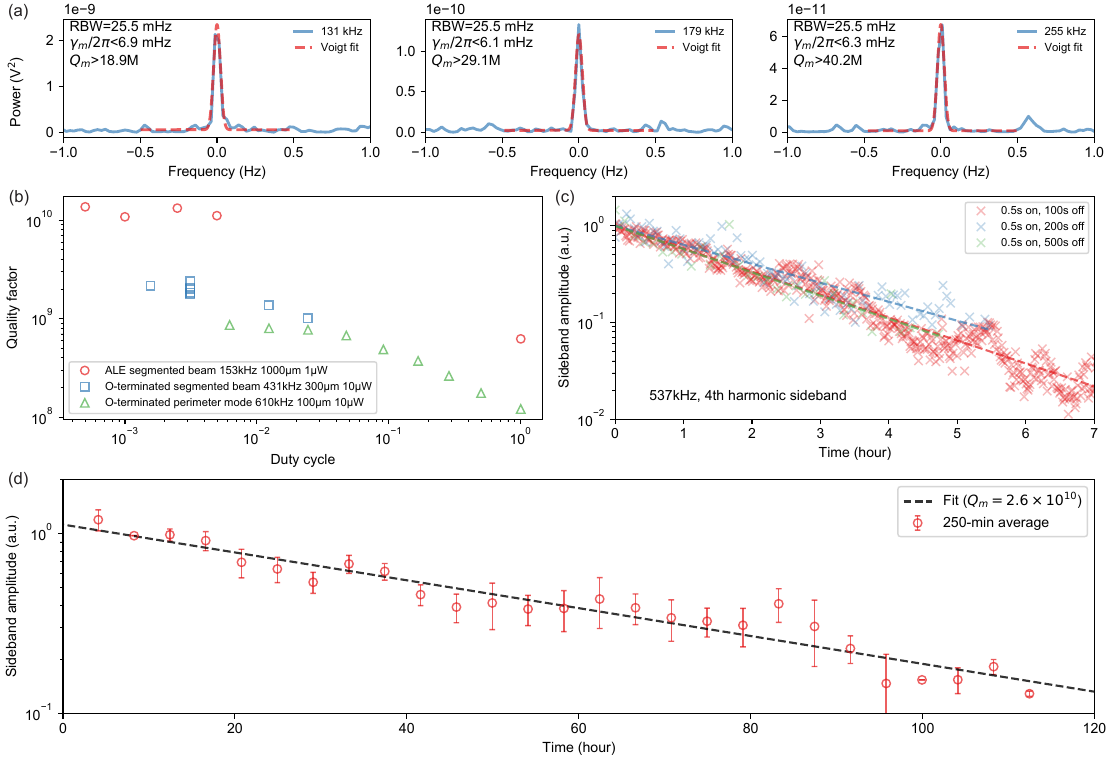}  
    \caption{\textbf{(a)} Spectral measurements of a perimeter mode device operating in a dilution refrigerator, showing the fundamental, second order, and the soft-clamped modes at different frequencies. These measurements show that the upper bound of the dephasing rate is significantly lower than the thermal decoherence rate of these resonators, demonstrating lifetime-limited performances. Voigt function is used to upper-bound the unresolved mechanical linewidth $\gamma_m$ with 95\% confidence interval, using a Blackman-Harris window function with resolution bandwidth (RBW) of \SI{25.5}{mHz}. \textbf{(b)} Quality factor measurement as a function of stroboscopic duty cycles, for three devices with varying structure sizes and probe powers, showing that the optical backaction effect is negligible for duty cycles $<\SI{e-2}{}$. \textbf{(c)} Stroboscopic ring down measurement of the ALE segmented beam device at \SI{12}{mK}, with varying duty cycles. The 4th harmonic sideband is used to accelerate the ring-down measurement. Even so, the measurement records show a significant long-term drift of optical alignment to the device inside the dilution refrigerator when the time scale approaches hours. \textbf{(d)} Ring-down measurement of a 42-kHz segmented beam device, averaged using multiple stroboscopic ring-down traces to suppress drift-induced errors. The x-axis shows the first-harmonic sideband decay time, projected from higher-harmonic records. }
  \label{fig:SI_Fig2_2}
\end{figure*}

\begin{table}[t]
\begin{center}
\caption{Comparison of intrinsic acoustic quality factors $Q_{\mathrm{int}}$ for various high-aspect-ratio nanomechanical material platforms, highlighting differences between amorphous and crystalline phases and their respective thicknesses.}
\label{table:1}
\begin{tabular}{ | c | c | c| c | } 
  \hline
  Material& Phase & $Q_{\mathrm{int}}$ & h (nm)\\ 
  \hline\hline
  SiN~\cite{Engelsen2024} & Amorphous & 2.5e3 & 20 \\ 
  \hline
  4K-Si~\cite{Beccari2022} & Crystalline & 8.0e3 & 14 \\ 
  \hline
  InGaP~\cite{Manjeshwar2023} & Crystalline & 8.1e3 & 73 \\ 
  \hline
  AlN~\cite{Ciers24} & Crystalline & 8.0e3 & 290 \\ 
  \hline
  SiC~\cite{Romero20} & 3-C & 1.2e3 & 100 \\ 
  \hline
  SiCOI~\cite{Sementilli2025} & 4-H & 1.5e4 & 100 \\ 
  \hline
  SiC~\cite{Hochreiter25} & 4-H & 2.8e4 & 100 \\ 
  \hline
  SiC~\cite{Xu23} & Amorphous & 5.2e3 & 100 \\ 
  \hline
  \textbf{5K-Diamond} & \textbf{Crystalline} & \shortstack{\textbf{1.3e6} (e-grade) \\ \textbf{5.4e5} (o-grade)} & \textbf{100} \\
  \hline
  \textbf{12mK-Diamond} & \textbf{Crystalline} & \shortstack{\textbf{70e6} (e-grade) \\ \textbf{5e5} (o-grade)} & \textbf{100} \\
  \hline
  \textbf{12mK-Diamond} & \textbf{Crystalline} & $>$\textbf{100e6} (e-ALE) & \textbf{150} \\
  \hline
\end{tabular}
\end{center}
\end{table}

Mechanical quality factors of the diamond nanomechanical devices were characterized using a homodyne/heterodyne Michelson interferometer setup (see Fig.~\ref{fig:SI_Fig2}(a)). A \SI{730}{nm} external-cavity diode laser (Toptica ECDL) and a TiSa laser provided approximately \SI{1}{mW} optical power, from which less than \SI{10}{\micro W} was directed to the sample. The laser beam was focused onto the diamond resonator through a 4f-confocal optical system employing a high-NA (0.9) objective (1-mm work distance), yielding a 500-nm spot size comparable to the device width. Beam positioning was precisely controlled using a fast steering mirror. The reflected signal from the device was recombined with a local oscillator frequency-shifted by a small frequency offset via a fiber-coupled double-acousto-optic modulator (dAOM), enabling heterodyne detection with a balanced photodetector. This method effectively suppressed the need for active optical path-length stabilization during ringdown measurements and stroboscopic measurements. For spectral linewidth measurement, the interferometer phase is stabilized using the dAOM with continuous measurements.

Samples were affixed to a holder using fast-drying silver paste and a miniature piezoelectric actuator (piezo) to the backside of the holder using epoxy. This piezo actuator allowed acceleration-based excitation of mechanical resonances. The sample holder was integrated into an Attodry cryostat capable of stable temperature sweeps from \SI{5}{K} to \SI{295}{K}, with measured external vacuum levels between \SI{2e-5}{mbar} and \SI{1e-6}{mbar}. For the dilution refrigerator measurement, the temperature is stabilized by heaters at the mixing plate using PID-feedback, allowing access to temperatures between \SI{10}{mK} to \SI{3}{K}. Actual pressure at the sample stage was expected to be significantly lower during cryogenic operation, and is estimated to be on the order of \SI{e-13}{mbar}. To reach the lowest operation temperature for the dilution refrigerator, since a free-space optical interferometer requires optical windows, we need to eliminate excess heat from the external black body radiation at room temperature. Even though the silica windows are not transparent to most of the black body radiation spectrum, residual radiation for wavelength $<\SI{4}{\micro\meter}$ can still reach the mixing plate, causing about 100uW of heating. We find that by minimizing the opening of the window at the 50-K stage using aluminum foil, and turning off all the lighting in the lab, the base temperature is significantly improved from \SI{26}{mK} to \SI{10}{mK}. We expect a cryo-switchable beam block could further reduce the base temperature to \SI{7}{mK} while allowing free-space optical access. 

Mechanical resonances were excited by sweeping the frequency of the piezo drive across resonance or by shaking the nearby attocube, while the resulting device displacement was tracked in real-time using a phase-locked loop (PLL) for continuous measurement, or a fixed bandwidth lock-in amplification at the sideband frequency. Resonance ring-down data obtained from the PLL provided both amplitude and frequency time-series, while the lock-in data only provided the amplitude time-series, from which mechanical quality factors were directly extracted. Laser power was systematically varied to verify that measured quality factors were unaffected by photothermal amplification or damping (example see Fig.~\ref{fig:SI_Fig2}(c)). For measurement down to mK temperature, we use stroboscopic probing using an automated beam shutter, with measurement results and duty cycle dependence of varying device dimensions and probe power in Fig.~\ref{fig:SI_Fig2_2} (b,c), and the best device shown in Fig.~\ref{fig:SI_Fig2_2} (d). 
To show that the device performances are limited by the lifetime, we also perform spectral linewidth integration using a homodyne measurement. The spectral-domain measurement of quality factors is more challenging, due to nonlinear effects (e.g. the Duffing nonlinearity), long-term temperature drifts, and the phase noise of the interferometer loop. To minimize the effect of temperature drifts, we select a resolution bandwidth of our instrument to be \SI{25}{mHz} for the spectral measurement and maximize the phase stabilization gain using the acousto-optic modulator. We average the Brownian motion of the oscillators to avoid the Duffing nonlinearity-induced frequency drifts, with results shown in Fig.~\ref{fig:SI_Fig2_2}(a). Considering the Gaussian-broadening of the Blackman-Harris filter employed, we upper-bound the underlying mechanical linewidth with 95\% confidence interval, showing spectral-domain quality factors approaching those measured using the ring-down approach. This level of dephasing (\SI{7}{mHz} upper-bound) is smaller than the best thermal decoherence rate of our system $\Gamma_{\mathrm{th}}/2\pi = \SI{9}{mHz}$, showing lifetime-limited performances.

\section{Van der Waals self-assembly}\label{app:vdW}

\begin{figure}[t]
    \includegraphics[width = 0.495\textwidth, page = 1]{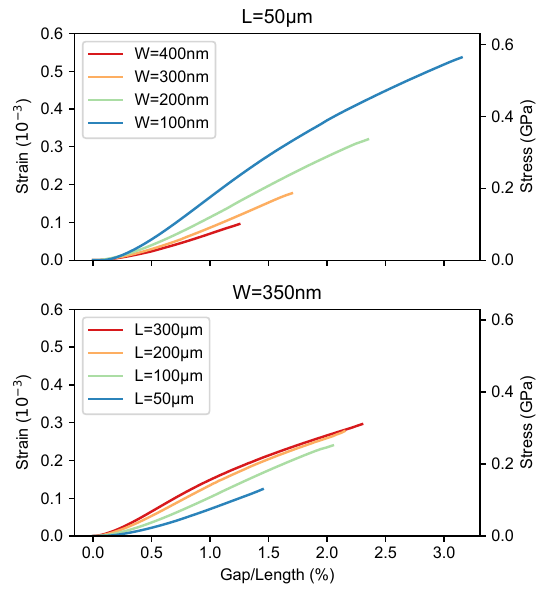}  
    \caption{COMSOL simulation of vdW adhesion-induced tensile strain between pairs of uniform diamond beams, assuming an adhesion energy of $\Delta\gamma = \SI{80}{mJ/m^2}$. Beam width ($W$) is varied at fixed length ($L=\SI{50}{\micro\meter}$), and the beam length is varied at fixed width ($W=\SI{350}{nm}$). The resulting tensile strain is plotted against the initial separation gap, with curves terminating at complete decohesion of the two beams.   }
  \label{fig:SI_Fig3}
\end{figure}

\begin{figure*}[t]
    \includegraphics[width = 0.9\textwidth, page = 1]{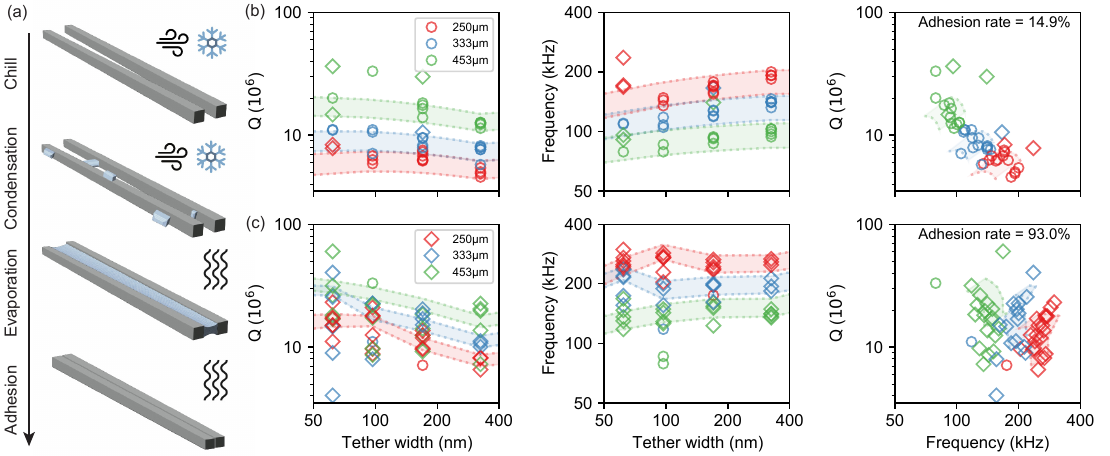}  
    \caption{\textbf{(a)} Schematic of the liquid-assisted vdW self-assembly procedure. The devices are initially cryogenically cooled below the freezing point inside the cryostat and then briefly exposed to ambient air to induce controlled condensation on device surfaces. Condensation is halted by sealing the cryostat and initiating vacuum pumping. As the condensed water evaporates completely, devices self-assemble via vdW adhesion. \textbf{(b)} Mechanical quality factors and resonant frequencies measured at \SI{5}{K} before self-assembly, plotted versus tether width. Devices unintentionally adhered after critical-point drying (15\%) are indicated by diamond symbols, while non-adhered devices are represented by circles. \textbf{(c)} Post-self-assembly mechanical characterization at \SI{5}{K}, showing a high self-assembly yield (93\%). Shaded regions represent COMSOL simulation predictions for device frequency and mechanical quality factors, assuming surface adhesion energy between $\Delta \gamma = \SI{30}{}-\SI{80}{mJ/m^2}$ and the intrinsic mechanical quality factor $Q_{\mathrm{int}}=\SI{3e5}{}$. The discontinuity observed in simulation results corresponds to the decohesion phenomenon detailed in Fig.~\ref{fig:SI_Fig5}. Notably, the device achieving the highest quality factor \SI{60e6}{} at \SI{453}{\micro\meter} fully adheres beyond simulation expectations, surpassing predicted partial decohesion behavior for this device length. }
  \label{fig:SI_Fig4}
\end{figure*}

\begin{figure*}[t]
    \includegraphics[width = 1\textwidth, page = 1]{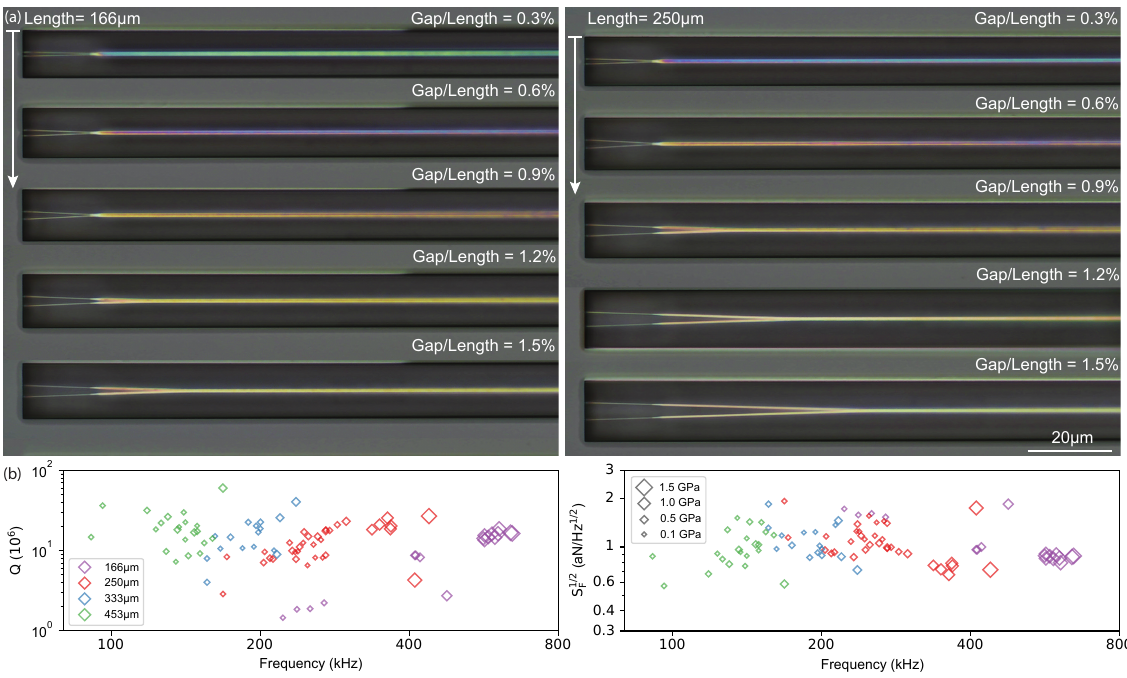}  
    \caption{\textbf{(a)} Optical microscope images of vdW self-assembled clamp-tapered diamond beams at various initial separation gaps, highlighting different decohesion thresholds in the wider beam regions. For beam lengths of \SI{166}{\micro\meter}, decohesion occurs at a gap-to-length ratio of 1.2\%; for beams \SI{250}{\micro\meter} long, this occurs at a gap-to-length ratio of 0.9\%. These observations align closely with stress measurements presented in the main text Fig.~2(g), where the decohesion events manifest as distinct jumps in achievable tensile strain. \textbf{(b)} Acoustic quality factors measured at \SI{5}{K}, force sensitivity and corresponding resonance frequencies are also summarized, showing enhancement from stress (encoded in marker size).}
  \label{fig:SI_Fig5}
\end{figure*}

\begin{figure}[t]
    \includegraphics[width = 0.495\textwidth, page = 1]{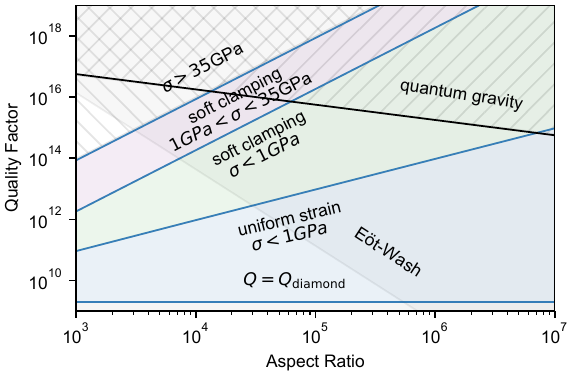}  
    \caption{Estimate of quality factor of diamond nanomechanics accessible through dissipation dilution and mK-operation. Two parameter regimes are highlighted where low-energy physics can be probed using cryogenic mechanical oscillators. First is the current sensitivity bound set by the Eöt-Wash torsional balance experiment for the detection of ultra-light dark matter for one-day averaging using diamond nanomechanical systems. Second is the threshold where the non-classical nature of gravity can be examined through gravitationally induced entanglement between two gravitationally interacting solid-state masses. }
  \label{fig:SI_Fig6}
\end{figure}

For diamond, first-principles calculations provide a reliable estimate of the ideal vdW surface adhesion energy:
\[\Delta \gamma_\text{ideal}=A_H/12\pi z_0^2 \approx \SI{80}{mJ/m^2}\] 
where $A_H = \SI{2.9e-19}{J}$ is the Hamaker constant determined by Lifshitz theory~\cite{BERGSTROM1997}, and $z_0 = \SI{0.3}{nm}$ is the equilibrium vdW separation~\cite{SUN2015,BERGSTROM1997}. This level of surface energy is modest due to the nonpolar nature of the diamond lattice. Due to the slight surface graphitization observed from X-ray photoelectron spectroscopy (XPS) after oxygen plasma etching, we estimate a modest adjustment of the Hamaker constant to approximately $A_H = \SI{2.6e-19}{J}$ (amorphous graphite layer), introducing a 10\% uncertainty in the actual surface adhesion potential.

Realistically, nanoscale surface roughness significantly reduces the effective vdW adhesion potential through spatial averaging, described by the Rabinovich–Rumpf factor of a rigid body~\cite{RABINOVICH2000}:
\[f(R,\lambda) = \frac{8}{3}\left(\frac{z_0}{z_0+R}\right)^2\left(1+\frac{1}{(1+1.82\frac{R}{z_0})^2}\right)\]
where $R$ denotes surface roughness. Here, the expression is simplified in the large aperiodic length limit (\SI{60}{\nm} measured from AFM). The model assumes the interface is made up of a few widely spaced asperity summits rising from an otherwise flat valley floor, so adhesion is governed mainly by contacts between those summits and the opposing flat areas—an assumption that agrees well with the summit-valley pattern observed in our AFM measurements. To accurately quantify the effective adhesion potential of our fabricated devices, we employed a double-beam geometry, with two adjacent nanomechanical beams initially separated by a controlled gap (main text Fig.~2). Upon self-assembly through vdW adhesion, the adhered length depends on a competition between adhesion and mechanical tearing forces originating from the device clamps. Comparing experimentally measured adhesion lengths against finite-element (FEM) simulations allowed us to precisely extract the effective vdW potential on a device-by-device basis. This approach provides a universal, non-destructive method for evaluating sidewall roughness at the nanoscale, independent of device thickness.

Improving surface smoothness enables approaching the ideal adhesion potential and thus maximizing tensile stress, which scales approximately as $\sigma\propto \Delta\gamma^{1/3}$. From the numerical simulation shown in Fig.~\ref{fig:SI_Fig3}, the tensile stress favors narrow beam widths ($\sigma\propto W^{-2/3}$) and high aspect ratios. Narrow beam widths significantly enhance achievable stresses by effectively reducing the mechanical tearing force. We therefore introduced tapered clamp regions, narrowing down to approximately \SI{50}{nm} at device ends to minimize mechanical tearing and enhance tensile strain, while maintaining wider central regions suitable for high-efficiency free-space optical displacement readout.

Realizing robust self-assembly in high-aspect-ratio structures has previously been challenging, with earlier techniques primarily limited to relatively short devices~\cite{Ashiba11} ($\SI{<50}{\micro\meter}$). To address this, we developed a selective liquid-assisted vdW self-assembly process based on picoliter-scale water condensation, shown in Fig.~\ref{fig:SI_Fig4}. During the assembly process, the capillary force, which originates from the polar nature of the liquid molecules, initiates the self-assembly by pulling the beams inward, while the vdW potential fixes the final geometry of the assembly. Hence, the term ``liquid-assisted vdW self-assembly" is used to describe our approach. We first cool down the device below the freezing point (-5 to 0 degrees Celsius) and vent with humid air. The liquid deposition process is visually monitored using the CMOS camera image and subsequently terminated when micron-scale icicles and/or droplets appear. Pumping and heating are followed for water evaporation. Unlike conventional liquid immersion methods that risk undesired substrate adhesion, our approach controls condensation at the suspended beam interfaces, resulting in a highly reproducible (\SI{93}{\%}) self-assembly success rate for devices up to \SI{500}{\micro m} long, thus compatible with scalable mass fabrication. We have also found that flashing the sample (at room temperature) with molecular-grade water steam is equally effective, terminating the flushing by looking for faint condensation texture indicative of micron-scale droplets. 

Experimentally, tensile stresses as high as \SI{1.3}{GPa} were realized in clamp-tapered designs, closely matching FEM predictions for effective vdW potentials between 30-\SI{80}{mJ/m^2} (main text Fig.~2(g)). Stress enhancement at narrower clamp widths arises from a combination of factors, including improved vdW adhesion due to reduced decohesion force and stress redistribution from the width modulation; the latter can contribute up to a twofold enhancement, factoring in thickness variations inherent in quasi-isotropic etching. For devices with systematically varied separation gaps, both measurements and simulations consistently showed abrupt stress changes corresponding to sudden decohesion of partially adhered regions. This behavior was further corroborated through direct optical microscope observations (Fig.\ref{fig:SI_Fig5}), visually confirming decohesion events at predicted gap thresholds. Note that with partial decohesion, the attainable tensile stress is no longer straightforwardly determined by the designed geometry, but also relies on the vdW potential holding the beams together.

Because vdW strain-engineering is fundamentally material-agnostic, it can be extended to a wide class of crystalline platforms that lack built-in stress~\cite{Hochreiter25,Shandilya2019,Machielse19}. Softer single-crystal materials such as silicon and GaAs with comparable surface potentials could support even larger tensile strains, while the GPa-level strain demonstrated here in diamond is already sufficient to tune the electronic structure and operating temperature of strain-sensitive defects~\cite{Meesala18,Stas22}. 

\subsection*{Performance scaling of vdW structures}
Remarkably, intrinsic mechanical quality factors $Q_\text{int}= \SI{3e5}{}$ measured in double-beam optical-grade diamond structures (\SI{100}{nm} thickness, Fig.~\ref{fig:SI_Fig4}, Fig.~\ref{fig:SI_Fig5}), corrected for dissipation dilution based on main text Fig.~2, closely matched expectations derived from independent surface-loss characterization (main text Fig.~3(c)), after considering the increased surface-volume ratio around 50\%.  The absence of significant additional mechanical losses due to the vdW adhesion interface can be understood by the high-quality diamond surfaces and the symmetric alignment of the adhesion plane, minimizing dynamic interface distortions during mechanical oscillations. Our optimized vdW self-assembly therefore demonstrates both a practical solution for substantial tensile strain generation and a versatile methodology applicable broadly to other crystalline nanoscale platforms.

Given the strain-width relation $\sigma\propto \Delta\gamma^{1/3}W^{-2/3}$, we anticipate that devices with even smaller tether widths, down to \SI{50}{nm}, and a gap-length ratio up to 2.5\% could achieve a ten-fold increase in available strain. This will require different read-out mechanisms (e.g. integrated optical cavity based ones), however, since the free-space interferometric readout currently in use requires widths to be larger than \SI{250}{nm}. Furthermore, large strain enhancement can be achieved locally by spatially varying the cross-section of the beam. This would enable the study of diamond color centers under extreme strain conditions. 

Further improvements will require an increase in the aspect ratio, currently in the $10^3$-$10^4$ range, which is smaller than $10^6$ already demonstrated in SiN~\cite{Engelsen2024} and strained-Si~\cite{Beccari2022} platforms. The aspect ratio is especially important for soft-clamping structures due to their quadratic scaling. The quality factor is improved by $Q = D_Q Q_{\mathrm{int}}$ with dilution factor~\cite{Fedorov19}:
\begin{equation}
    D_Q = 1+\frac{1}{\alpha \lambda + \beta \lambda^2}, \quad \lambda = \sqrt{\frac{1}{12\epsilon}}\,\frac{1}{r},
\end{equation}
where $r=L/h$ is the aspect ratio, $\epsilon$ the tensile strain, and $\alpha$ and $\beta$ geometry-dependent form factors describing clamping and distributed losses. To reach these high aspect ratio values in diamond will require advancements in diamond nanofabrication strategies. Quasi-isotropic undercut techniques currently utilized result in resonator-bulk diamond separation of only \SI{5}{\micro\meter}, which limits how long devices can be suspended without stiction issues. For comparison, state-of-the-art SiN devices typically employ $>$\SI{20}{\micro\meter} of separation from the substrate to avoid stiction. These large separations could be achieved in diamond by further optimizing the quasi-isotropic etching process or by using recently emerged thin diamond films~\cite{Guo2021}. The second technical difficulty is the trade-off between the aspect ratio and the device statistics, as diamond substrates typically have a very small suitable area of \SI{2}{mm}$\times$\SI{2}{mm}, therefore limiting the number of devices that can be fabricated. We are aware that industry vendors, such as DiamondFoundary, have been supplying the supercomputing industry with single-crystal diamond plates at \SI{5}{cm} scales with modest costs, and the acquisition of those samples could lead to devices with aspect ratios approaching millions. 

The $Qf$ product, frequently used as one of the figures of merit in the field of cavity optomechanics, of our devices is at best around \SI{2e15}{Hz}. When comparing to the best value in diamond optomechanical crystal cavities (\SI{e16}{Hz})~\cite{Hyunseok25}, the penalty from the low mechanical frequency causes a big gap to the state-of-the-art devices. However, for sensing applications, such as quantum sensing~\cite{Bertone2018}, tests of quantum gravity~\cite{Marletto17}, and interfacing color centers~\cite{Fung24}, $Q/f$ is instead the better figure of merit, which prefers low frequency oscillators. This is evident by the force sensitivity $S_F=4k_BTm_\mathrm{eff}\Omega_m/Q$. This point will be more evident in the discussion that follows.  

\section{Projection of high-aspect-ratio performances and applications in gravitational physics}
\label{app:cryo}
We estimate the optimal performance of strained single-crystal diamond nanomechanical oscillators operating at millikelvin (mK) temperatures and much higher aspect ratios. Assuming our measured material quality factor trend persists at even lower temperature, and setting aside other factors such as radiation loss, gas damping, and emerging loss channels, we project the achievable intrinsic quality factor $\sim10^9$ at \SI{1}{mK}, as shown in Fig.~\ref{fig:SI_Fig6}. 

Beyond reaching quality factors on par with optical clock transitions, these ultra-coherent oscillators are particularly promising for sensing weak forces in low-energy physics. Two notable applications include the detection of ultra-light dark matter and tests of the quantum nature of gravity—both of which demand extremely high acceleration sensitivity.

In the context of dark matter detection, one proposed method~\cite{Manley21} involves a pair of mirrors composed of materials with different baryon-lepton (B–L) ratios. Due to coupling with a uniform background of hypothetical vector dark matter (“dark photons”), these mirrors experience a differential force, resulting in a measurable relative displacement. The force can be optically read out, and the resulting data is used to place bounds on the coupling strength $g$ of the dark matter. The current best bound is set by the Eöt-Wash torsion balance experiment at approximately $g\approx \SI{e-22}{}$. For a resonant force sensor, the projected bound for $g$ in the long-integration-time limit $\tau\gg\tau_\mathrm{DM}$ ($\tau_\mathrm{DM}$ the coherence time of the dark matter) is given by
\[g\sim \frac{1}{f_{12} a_0}\sqrt{\frac{4k_BT\Omega_\mathrm{DM}^2}{mQQ_\mathrm{DM}}}(\tau_\mathrm{DM}/\tau)^{1/4},\] where $a_0=\SI{3.7e11}{m/s^2}$, $Q_\mathrm{DM}\sim\SI{e6}{}$ is the effective quality factor of the dark matter, and $f_{12}=0.05$ accounts for the B–L ratio difference between the diamond and beryllium mirrors. The corresponding intrinsic quality factor required to beat the Eöt-Wash bound scales as $Q_\mathrm{int}\propto h/L^5$, where $h$ is the thickness of the resonator, strongly favoring large aspect ratios. The required performance threshold by one-day integration is indicated by the light gray region in Fig.~\ref{fig:SI_Fig6}.

Similarly, in the proposed scheme for testing the quantum nature of gravity via gravitationally induced transparency (GIT)~\cite{mari2025}, two mechanical resonators are brought into proximity such that their gravitational interaction dominates over the interaction with the environment bath (other forces eliminated by electrical shielding). For two semi-infinite string-like objects, the maximum gravitational coupling rate is given by: 
\[\lambda = Gm/Ld^2\Omega_m \approx \Omega_G^2/\Omega_m,\] 
where $G$ is the gravitational constant, $m$ is the mass of the oscillator, $L$ is the resonator length, $d$ is the separation gap between the two, $\Omega_m$ is the mechanical frequency, and $\Omega_G = \sqrt{\pi G \rho} = \SI{8.6e-4}{Hz}$ is a gravitational material constant. 

Since these are low-frequency oscillators, it is useful to define the environment bath frequency at unity phonon occupation $\Omega_T = k_BT/\hbar$. 
In the case where a quantum signal can be transferred from one mechanical oscillator to the other, and the channel is not entanglement-breaking, then gravity must have a quantum nature (model independent). Mapping the system to a quantum thermal attenuator channel, the criteria for reaching this threshold are quantified as $\Omega_G^2/\Gamma_m\Omega_T\geq 1$, and equivalently the threshold for oscillator quality factor $Q\geq \Omega_m \Omega_T/\Omega_G^2$. 

For soft-clamped resonators, the intrinsic quality factor needed to reach this regime scales as $Q_\mathrm{int}\propto h^2/L^3$, again favoring high aspect ratios. This gravitational quantum test threshold is highlighted as the patched region in Fig.~\ref{fig:SI_Fig6}, assuming a 1-mg mass-loaded resonator.

We note that these projections are conservative. Cryogenic platforms based on nuclear demagnetization have demonstrated temperatures below \SI{100}{\micro\kelvin}~\cite{Yan21}, and further improvements—such as mass-loading in soft-clamped geometries—are expected to substantially lower the quality factor thresholds for both ultra-light dark matter detection and quantum gravity tests. Note that our current system parameters demonstrated in this manuscript are about $10^{-10}$ away from the threshold, and to our knowledge, one of the closest value to date~\cite{Bose25}, with the other system types such as the delocalized nanoparticles~\cite{Hofer23,Steiner25} and atom interferometers~\cite{Kovachy2015} well below $10^{-10}$. We emphasize that these are soft estimates and depend on the assumptions of the implementation of Faraday shielding for different types of systems. 

Lastly, we discuss the application in resolving the quantum-classical boundary problem, which requires additional mass loading to reach the required sensitivity target. The continuous spontaneous localization (CSL) model~\cite{Ghirardi90} modifies Schrödinger dynamics by adding a universal, stochastic localization with rate $\lambda\sim 10^{-16}\,\mathrm{s^{-1}}$ and length scale $r_C\sim10^{-7}\,\mathrm{m}$ (the GRW point~\cite{Ghirardi86}).
This is chosen so that microscopic systems behave essentially like ordinary quantum mechanical systems, whereas macroscopic superpositions collapse extremely fast, to resolve the measurement problem. Note that other parameterization choices also exist~\cite{Schrinski23}.

Because of that, if experiments rule out that point (no added noise at that level), then the original GRW model with those parameters is falsified. The current bound $\lambda \sim 10^{-11}$ at $r_C\sim10^{-5}\,\mathrm{m}$ is set by a layered cantilever experiment~\cite{Vinante2019} at cryogenic temperature. As the bound $\lambda \propto T\Omega_m/Q m_\mathrm{eff}$, we project that with a high-aspect ratio $r=10^6$, hard-clamped structure with mass-loading to $m_\mathrm{eff}=\SI{1}{\micro\gram}$ and a quality factor of $Q=\SI{e14}{}$, we can reach $\lambda\sim \SI{e-17}{}$ and $r_C\sim\SI{e-7}{m}$. Even though this is just slightly beyond the GRW point, such an experiment will probe parameter regions that greatly extend the current quantum-classical boundary, and could set very tight bounds on the existing CSL models. For more near-term bounds, such as a test for the Diosi-Penrose model~\cite{Penrose2014} describing the collapse due to the gravitational self-energy, our system is already at least two orders of magnitude beyond the threshold, where a follow-up decoherence rate measurement can be performed to implement the test directly.  

\section{TLS theory and experiment}
\newcommand{\kB}{k_{\mathrm{B}}}
\newcommand{\hbarc}{\hbar}
\newcommand{\w}{\omega}
\newcommand{\Ea}{{E_{\mathrm{a}}}}
\newcommand{\El}{E}
\newcommand{\Dl}{\Delta}
\newcommand{\Dlt}{{\Delta_0}}
\newcommand{\rhoM}{\rho}
\newcommand{\vl}{v_l}
\newcommand{\vt}{v_t}
In the following, we introduce two TLS models: the Arrhenius thermal relaxation model that describes hopping between two nearly symmetric configurations separated by an energy barrier of $E_a$; and the quantum TLS relaxation model that describes hopping between two quantized energy levels separated by the energy difference $E_a$. Here we call both energy quantities the activation energy, since both models share very similar physics, but we will also describe their subtle differences at the end of the section. Note that these are relaxational models that lead to off-resonant energy losses. Differing from resonant energy loss to TLS, relaxational processes describe the population of a TLS attempting to relax (through hopping or damping) towards an equilibrium point, which is being modulated due to strain coupling to the resonator motion. Because there is no resonant phonon absorption or emission, there is no TLS saturation effect observed in the resonant case.
 
\subsection{Arrhenius thermal relaxation}
Arrhenius thermal relaxation describes acoustic losses from strain-coupling to a single defect that hops between two configurational minima with barrier energy $\Ea$. The dynamical strain modulates the minima energies, therefore periodically shifts the thermal equilibrium population. Since thermal equilibrium is essentially a dissipative process by coupling to a thermal environment, this mechanism leads to a relaxational internal friction with characteristic dependence on the Arrhenius relaxation time $\tau$,
\begin{equation*}
\label{eq:Qinv_arr_single}
Q^{-1}_{\mathrm{A}}(T;\Ea,\w)\propto\frac{\w \,\tau(T;\Ea)}{1+\w^2 \tau^2(T;\Ea)}\,,
\end{equation*}
with $\tau$ following the canonical thermal activation scaling to temperature $T$,
\begin{equation}
\label{eq:tau_Arr}
\tau(T;\Ea)=\tau_0(\Ea)\,\exp\!\left(\frac{\Ea}{\kB T}\right).
\end{equation}
The factor $\w\tau/(1+\w^2\tau^2)$ is the Debye relaxational kernel. It peaks at temperature $T^*$ when $\w\tau(T^*)=1$. The Debye kernel follows naturally by considering a defect with internal variable $X$ (e.g.\ population imbalance between the two configurations) relaxing toward the equilibrium $X_{\text{eq}}(\varepsilon)$ with a relaxation time constant $\tau$:
\begin{equation*}
\dot X = -\frac{1}{\tau}\left[X - X_{\text{eq}}(\varepsilon)\right],\qquad
X_{\text{eq}} = X_0 + S\,\varepsilon .
\end{equation*}
Under a harmonic strain drive $\varepsilon=\varepsilon_0 e^{i\omega t}$ and coupling coefficient $S$, the frequency domain solution gives a complex modulus $M(\omega)=S/(1+i\omega\tau)$ and the loss angle
\begin{equation*}
Q^{-1} = \frac{-\mathrm{Im} M}{M_0} \simeq \left(\frac{S}{M_0}\right)\frac{\omega\tau}{1+\omega^2\tau^2}.
\end{equation*}
Mapping this to a defect that can occupy two configurations $\ket{1}$ and $\ket{2}$ whose energies differ by an asymmetry $2\delta$. Under an applied strain $\varepsilon$, the energy asymmetry shifts because each configuration has a different elastic dipole $\Lambda_i$:
\begin{equation*}
E_{\pm}(\varepsilon) = \pm(\delta + \gamma\,\varepsilon),
\qquad
\gamma = \frac{1}{2}(\Lambda_2-\Lambda_1)\,,
\end{equation*}
where $\gamma$ is the strain coupling constant, with units of energy per unit strain. Typical magnitudes are $\gamma \sim \SI{1}{eV}$ for covalent bonds.
For number density $n$ of identical defects, the partition function and free energy per volume are
\begin{equation*}
Z = 2\cosh\!\left(\frac{\delta+\gamma\varepsilon}{k_B T}\right),\qquad
F(\varepsilon) = -n k_B T \ln Z.
\end{equation*}
The defect contribution to the elastic modulus $\mathcal{S}$ (generated stress per unit strain) follows from the second derivative of $F$ with respect to strain,
\begin{align*}
\mathcal{S} = \frac{\partial^2 F}{\partial \varepsilon^2}\bigg|_{\varepsilon=0} = \frac{n\,\gamma^2}{k_B T}\,\text{sech}^2\!\left(\frac{\delta}{k_B T}\right).
\end{align*}
Plugging this result into our previous expression of $Q^{-1}$ yields
\begin{equation}
    Q^{-1}(\omega,T)= \frac{\mathrm{Im}[\mathcal{M}]}{M_0}= \frac{n\,\gamma^2}{\rho v^2\,k_B T}\,\text{sech}^2\!\left(\frac{\delta}{k_B T}\right)\,\frac{\omega\tau(T)}{1+[\omega\tau(T)]^2}.
\label{eq:arrhenius_core}
\end{equation}
where $M_0\simeq\rho v^2\sim E$ is the elastic (Young's) modulus of the resonator mode. In the following, we assume $\delta\ll k_BT$ for symmetric configurations.

\subsection*{Scaling of defect parameters to activation energy}
Since our measurement probes defects with activation energy spanning orders of magnitude, in the case of such a large distribution of defects, we can write the total loss as
\begin{gather}
     Q^{-1}(T) =\sum_i\int_0^\infty\!P_i(E_a)\,\frac{n_i\,\gamma_i(E_a)^2}{\rho v^2\,k_B T}\nonumber \\
     \times \frac{\omega\tau_{0,i} e^{E_a/k_B T}}{1+[\omega\tau_{0,i} e^{E_a/k_B T}]^2}\,dE_a,
\label{eq:Q_arrhenius_ens}
 \end{gather} 
Note that $P_i(E_a)$ is the defect distribution in $E_a$ for the same defect family $i$, whereas $n_i$ is supposed to capture the actual spatial density of the defect family. Due to spectral overlap, we can not fully separate these two distributions. We can, however, identify individual mode family contributions $P_i(E_a)n_i$ first based on the collective distribution $\sum_iP_i(E_a)n_i$ (shown in Fig.3 of the main text), then retrieve the defect density through $n_i = \int P_i(E_a)n_i dE_a$. Note that this is only possible by assuming a shared $\tau_0(E_a)$ across different mode families. 

In order to extract a faithful defect density $n_i$, we need to understand the basic scaling of defect parameters such as $\gamma$ and $\tau_0$ to the activation energy $E_a$. The thermal diffusion time constant $\tau_0$ is well described by the classical over-barrier hopping between two potential minima separated by a barrier of height $\Ea$ along a generalized coordinate $x$ with mass $m$. Near the left minimum $a$ and the saddle point $b$ (barrier top), we approximate them as a harmonic potential:
\begin{gather*}
V(x)\approx \frac12 m\omega_a^2 (x-x_a)^2 \quad (\text{well})\\
V(x)\approx \Ea - \frac12 m\omega_b^2 (x-x_b)^2 \quad (\text{barrier}).
\end{gather*}
Here $\omega_a$ is the small-amplitude oscillation frequency in the well and $\omega_b$ is the curvature magnitude at the saddle.

In Kramers theory~\cite{KRAMERS1940, Hanggi90}, the escape rate has the universal Arrhenius form with a prefactor that depends on dissipation:
\begin{align*}
\tau(T) = \tau_0 e^{\Ea/\kB T},
\end{align*}
We assume that in atomic defect systems, we are in the underdamped limit, where we have 
\begin{equation}
    \tau_0 = \frac{2\pi}{\Gamma}\frac{\omega_b}{\omega_a}\frac{k_BT}{E_a}
\end{equation}
where $\Gamma$ is the friction coefficient. Given that these are atomic-scale defects, the barrier curvature is typically associated with the well curvature $\omega_a\sim\omega_b$, depending only on the local atomic configurations, therefore only weakly on $E_a$. And since thermal activation typically requires $E_a\sim k_BT$, the only real dependence is $\tau_0\propto \Gamma^{-1}$. Since curvature typically characterizes energy-displacement coupling, we have $\Gamma\propto \omega_a$, therefore also only weakly depend on $E_a$. Given that there are many parameters depending weakly on $E_a$, in our fitting function, we set a weak energy dependence $\tau_0(E_a)=\tau_{00}(E_a/E_{\mathrm{ref}})^{-\mu}$, and find that $\mu=0$ gives the best fit, corroborated by our analysis. Since the defect at \SI{200}{K} allows us to retrieve both the $E_a$ and $\tau_0$, we use those as reference points for our studies. 

For the strain coupling constant $\gamma$, assuming the two energy minima are separated by $x_0$ with a small stiffness $k$, the potential barrier is $E_a\sim kx_0^2$ and the elastic dipole is $\Lambda = kx_0$. Since atomic scale stiffness is more or less a constant, we have $\gamma\propto \Lambda \propto \sqrt{E_a}$. In the fitting procedure, we set a weak dependence $\gamma(E_a) = \gamma_0(E_a/E_{\mathrm{ref}})^\alpha$, and find that $\alpha = 0.5$ indeed gives the best fit. 

We plot the characteristic temperature dependence of $Q$ in Fig.~\ref{fig:SI_TLS}(a) of different $\mu$ and $\alpha$ exponents, assuming an Arrhenius relaxation model with a uniformly distributed TLS in $E_a$ from an amorphous layer. The trend clearly indicates that $\alpha=0.5$ is the best scaling, whereas $\mu$ only shows weak dependence and is up to interpretations.  

\subsection*{Details of experimental characterization}

Due to the exceptionally high intrinsic quality factors achievable in diamond, identifying and isolating extrinsic loss channels is critical. Thermoelastic~\cite{Lifshitz00} and Akhiezer~\cite{Woodruff61} damping mechanisms were calculated to be negligible ($Q>$\SI{e8}{} at \SI{5}{K}) due to the high-aspect-ratio geometry, leaving gas damping and radiation losses at the device clamps as potential external loss sources. 

To robustly isolate intrinsic material losses, our first experimental realization utilized fractal soft-clamped resonators (Shown in Fig.~\ref{fig:SI_fabflow}), designed specifically to exponentially suppress acoustic radiation losses at the clamping regions by over two orders of magnitude compared to conventional string geometries of comparable dimensions. These devices do not have self-tensioning tethers, therefore not strained by much. The silver paste used underneath the chip lightly tensioned the chip for about \SI{25}{MPa}, which leads to a dilution factor of 28 for the electronic grade device and 21 for the optical grade device at 100-nm thickness, as shown in Fig.3(c) of the main text. Both electronic-grade and optical-grade diamond resonators underwent identical wet oxygen surface termination treatments prior to mechanical characterization. 

These chips were measured from room temperature to \SI{5}{K}, in an Attodry closed-loop cryostat. Temperature-dependent quality factor measurements revealed similar overall loss behaviors across both diamond grades. At room temperature, devices exhibited a consistent upper-bound quality factor of approximately \SI{e6}{}, independent of resonance frequency (Fig.~\ref{fig:SI_Fig2}(b)). This frequency independence effectively rules out gas damping~\cite{Martin08} as the dominant loss mechanism, given that gas damping would scale as: $Q_{\mathrm{gas}} = \SI{8.6e8}\cdot(f_{\mathrm{m}}/1[\text{MHz}])\cdot(h/20[\text{nm}])/(P/10^{-6}[\text{mbar}])$. 

Upon cooling below approximately \SI{270}{K}, this loss channel was eliminated, revealing a loss mechanism associated with thermally activated defects~\cite{Cannelli91} until \SI{50}{K}. Further analysis in the next section indicates their surface origins, corroborated by defect density calibration and X-ray photoelectron spectroscopy (XPS, Fig.~\ref{fig:SI_Fig1}). The measured activation energy (\SI{0.035}{eV}) and tunneling rate (\SI{9.1}{MHz}). 

Below approximately \SI{50}{K}, a distinct additional loss mechanism emerged. At \SI{50}{K}, electronic-grade diamond consistently demonstrated approximately twice the quality factor of optical-grade diamond. Detailed analysis (main text Fig.3(c)) indicates that this additional loss mechanism at cryogenic temperatures originates predominantly from surface defects introduced during oxygen plasma etching. These insights confirm surface treatment as a critical step in optimizing low-temperature mechanical coherence in diamond nanostructures. The fitted average intrinsic mechanical quality factor of both electronic and optical grades at \SI{4}{K} is compared to other material platforms in Table.~\ref{table:1}. To investigate the possible dependence of device performance on diamond crystal orientation, we have characterized devices fabricated along different crystal axes (22.5 deg difference between their orientations). We did not find noticeable differences in the intrinsic material quality factors.

After these chips, we fabricated the self-tensioning tether designs and moved the measurement from the Attodry cryo-system to a Bluefors dilution refrigerator, allowing us to access temperatures down to \SI{10}{mK}. By this larger temperature span, we realize that the loss plateau is the high-energy tail of a uniformly distributed amorphous defect layer that dominates the loss mechanism for temperatures below \SI{50}{K}. Besides the overall rising trend in the material quality factor that qualitatively matches that of the uniformly distributed amorphous defect layer, we observe oscillatory behaviors and discrepancies between different modes. Therefore, we proceed to extract information about these defects from these trends using the distributed Arrhenius loss model we derived earlier. 

\begin{figure}[t]
    \includegraphics[width = 0.495\textwidth, page = 1]{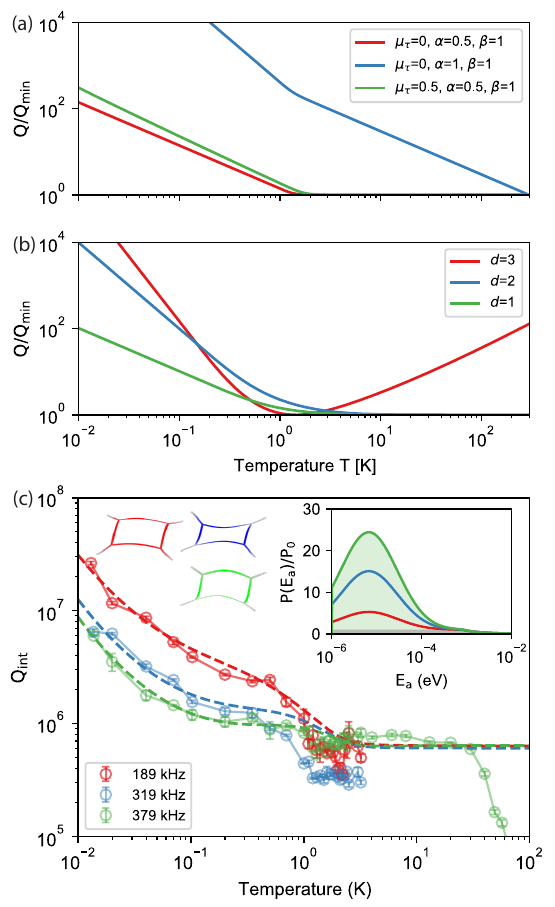}  
    \caption{\textbf{(a)} Temperature dependence trend of $Q$ for Arrhenius relaxation models with different scaling exponents, assuming a uniformly distributed TLS in $E_a$. \textbf{(b)} Temperature dependence trend of $Q$ for standard tunneling models with different phonon bath dimensions. \textbf{(c)} Temperature dependence of the material quality factors of the first three acoustic modes in a perimeter mode resonator, after hydrogen plasma followed by wet oxidation treatment. Dashed lines are collective fits of the Arrhenius relaxation model to the data, with the normalized defect density as a function of activation energy shown in the inset. }
  \label{fig:SI_TLS}
\end{figure}

In our fitting procedure, such as the one shown in Fig.3(e) and Fig.~\ref{fig:SI_TLS}(c), we first add a uniform background component to match the overall trend for $T<\SI{1}{K}$. Then, for every residue dip (we identified two), we add a new defect family with a log-normal distribution. We collectively fit the data from all three vibrational modes, with shared center $E_a$ and width $\Delta E$ for each defect family, but with independent peak densities to account for the mode-dependent coupling coefficient $\gamma$, which effectively modifies $P_i(E_a)n_i$. We found one defect family at $E_1=\SI{1.27e-4}{eV}$ with $\Delta E_1=\SI{1.58e-4}{eV}$ and another defect family at $E_2=\SI{1.25e-5}{eV}$ with $\Delta E_2 = \SI{1.56e-5}{eV}$.  $\Delta E$ is related to the log-normal width by $\sigma = \mathrm{arcsinh}(\Delta_E/2E)/\sqrt{2\mathrm{ln}2}$. 

For the defect densities shown in Fig.3(d) of the main text (oxygen plasma finishing), assuming a strain coupling constant $\gamma_0 = \SI{1}{eV}/\text{strain}$ and $E_{\mathrm{ref}}=\SI{1}{eV}$, we can estimate the surface defect density around $0.5/(100nm)^2$ for both defect families at \SI{e-4}{eV} and \SI{e-5}{eV}, whereas the surface density for the TLS loss peak at \SI{35}{meV} around \SI{200}{K} is around one defect per surface atoms, assuming a specific $\gamma=\SI{60}{meV}$ for rotor strain sensitivity. The background amorphous TLS is about $1/(100nm)^2$ on the surface. Note that these defects couple differently to different mechanical modes, as discussed in the main text. Here, the reported values are based on the mechanical mode that couples the strongest. If we use those as a reference, we can extract the fractional coupling power $\gamma^2/\gamma_\mathrm{max}^2$ for different defect families and different modes. For the defect family at \SI{e-5}{eV}, we have $14\%$ for the 1st mode, $94\%$ for the 2nd mode, and $100\%$ for the 3rd mode. For the defect family at \SI{e-4}{eV}, we have $100\%$ for the 1st mode, $65\%$ for the 2nd mode, and $18\%$ for the 3rd mode.  

Further, we fabricated another chip with an identical design as a control group, but it had undergone hydrogen plasma treatment and wet oxidation on the surface using Piranha solution. The measurement result of a similar perimeter mode resonator is shown in Fig.~\ref{fig:SI_TLS}(c). We perform the same collective fits and found the same defect family at \SI{e-5}{eV}, together with a uniform background. We find that the fractional coupling power $\gamma^2/\gamma_\mathrm{max}^2$ did not change much for the defect family at \SI{e-5}{eV}, we have $18\%$ for the 1st mode, $60\%$ for the 2nd mode, and $100\%$ for the 3rd mode. We further observed that the background uniform defect density increased to $5/(100nm)^2$, the density for the defect family at \SI{e-5}{eV} increased to $9/(100nm)^2$, and the density for the defect family at \SI{e-4}{eV} is not visible anymore. 

\subsection*{Inferring defect properties and origins}
Based on these observations, we could say something about the origins of these defects. First, based on the fact that the defect at \SI{35}{meV} is highly occupying the surface, corroborated with XPS results, and based on DFT results of Ref.~\cite{Holder13}, we can safely conclude that this is the -OH rotor defects. Second, based on the strong increase in the density of the \SI{e-5}{eV} defect after the hydrogen plasma treatment, and the identical surface oxidization finish, we can conclude that this is a defect complex that lives in the subsurface layer, and is primarily caused by hydrogen physics. Further, given the selective coupling to the 3rd mode that has large torsional strain, we can say that this is a defect that interacts more strongly with shear strain than with normal strain. Third, hydrogen is known for its ease of infusion and softens the defects. The increase in uniform defect density after hydrogen treatment indicates the presence of existing amorphous defects at the subsurface level. Lastly, the fact that the defect at \SI{e-4}{eV} only exists right after oxygen plasma treatment and disappears after hydrogen plasma + piranha treatment indicates that this is a defect that originates from oxygen at the subsurface level as well. And mode selectivity shows that this defect interacts more strongly with normal strain modes rather than the shear strain. 

\subsection{Standard tunneling model}
Differing from a configurational change, mechanical oscillators can also couple to quantum TLS. This is best described by the standard tunneling model, which assumes a uniformly distributed quantum TLS at low frequencies. In the following, we analyze the resulting relaxational damping rate scaling as a function of temperature, and deduce whether it is possible that our loss can also be explained by coupling to these systems.

Consider an atom (or small group) in a double-well potential with classical coordinate $x$. Projecting onto the two lowest localized states in left/right wells $\{\ket{L},\ket{R}\}$ yields the TLS Hamiltonian
\begin{gather*}
H_{\text{TLS}} = \frac{1}{2}\,\Delta\,\sigma_z + \frac{1}{2}\,\Delta_0\,\sigma_x ,\\
E=\sqrt{\Delta^2+\Delta_0^2},
\end{gather*}
where $\Delta$ is the asymmetry (energy difference between wells at zero strain) and $\Delta_0$ is the tunneling matrix element, and the eigenlevel spacing is $E$. 

In the energy eigenbasis, the strain couples longitudinally and transversely as
\begin{equation*}
H_{\text{int}} = -\sum_{ij}\gamma_{ij}\,\varepsilon_{ij} \,\sigma_z
= -\gamma_\parallel\,\varepsilon\,\tilde{\sigma}_z - \gamma_\perp\,\varepsilon\,\tilde{\sigma}_x,
\end{equation*}
where $\gamma_\parallel = \gamma \cos\theta$, $\gamma_\perp=\gamma\sin\theta$ with $\tan\theta=\Delta_0/\Delta$.
$\gamma_\perp$ drives level flips (energy exchange with phonons), $\gamma_\parallel$ modulates the gap (dispersive).

Decompose the strain field into phonon modes (longitudinal $\ell$, two transverse $t$) with strain $\varepsilon_{ij}=\tfrac{1}{2}(\partial_i u_j+\partial_j u_i)$. The interaction term $-\gamma_{ij}\varepsilon_{ij}\,\tilde{\sigma}_x$ drives transitions between TLS eigenstates with matrix element $\propto \gamma_\perp$. Using Fermi's golden rule for one-phonon emission/absorption at energy $E$ and summing different polarizations gives the TLS relaxation rate in the STM
\begin{equation*}\label{eq:tau_inv}
\tau^{-1}(E,\Delta_0;T) = \left(\frac{\Delta_0}{E}\right)^2 \frac{E^3}{2\pi \rho \hbar^4}
\left( \frac{\gamma_\ell^2}{v_\ell^5} + 2\frac{\gamma_t^2}{v_t^5} \right)\coth\!\left(\frac{E}{2k_BT}\right).
\end{equation*}
Here $\rho$ is mass density, $v_{\ell,t}$ are sound speeds, and the $E^3$ arises from the 3D phonon density of states and strain normalization. For a general d-dimensional system, the relaxation pathways change their densities. For acoustic phonons with linear dispersion $\omega_q = v|q|$ in spatial dimension $d$, the phonon density-of-state per unit volume scales as
\begin{equation*}
D_d(\omega)\ \propto\ \omega^{d-1}.
\end{equation*}
The strain amplitude of a phonon mode contributes an additional factor $\propto \sqrt{\omega}$, so that the Fermi–Golden–Rule one-phonon relaxation rate of a TLS with energy splitting $E$ scales as
\begin{gather}
\tau^{-1}(E,\Delta_0;T) = 
\left(\frac{\Delta_0}{E}\right)^{\!2}\, \frac{E^{d}}{\rho\hbar^{d+1}}\,\nonumber\\
\coth\!\left(\frac{E}{2k_B T}\right)
\Bigg(\frac{\gamma_\ell^2}{v_\ell^{\,d+2}} + (d-1)\frac{\gamma_t^2}{v_t^{\,d+2}}\Bigg).
\end{gather} 
Similar to the derivation in the Arrhenius model, we can derive the imaginary susceptibility of the TLS response
\begin{equation*}
    \chi''_{\mathrm{TLS}}(E,\Delta_0;T,\omega)
= \frac{\gamma^2}{\rho v^2 k_B T}
  \left(\frac{\Delta}{E}\right)^2
  \mathrm{sech}^2\!\left(\frac{E}{2k_B T}\right)
  \frac{\omega\tau}{1+\omega^2\tau^2},
\end{equation*}
with the relaxation-induced damping
\begin{gather}
    Q^{-1}_{\mathrm{STM}}(\omega,T) = \iint d\Delta\, d\Delta_0\, P(\Delta,\Delta_0)\, \frac{n\,\gamma^2}{\rho v^2\,k_B T} \left(\frac{\Delta}{E}\right)^2\nonumber\\
  \mathrm{sech}^2\!\left(\frac{E}{2k_B T}\right) \frac{\omega\tau}{1+\omega^2\tau^2}.
\end{gather}
which is almost identical to that of the Arrhenius model. There are only two subtle differences. The first is the scaling of $\tau_0$ with respect to the defect energy $E$ due to power scaling with different bath dimensions. Here, we use the standard STM distribution $P(\Dl,\Dlt)=P_0/\Dlt$ in glasses, a broad distribution uniform in $\Dl$ and log-uniform in $\Dlt$ within bounds. The $Q^{-1}(T)$ scaling is shown in Fig.~\ref{fig:SI_TLS}(b), where $d=1$ dimension matches our observation. At higher temperatures where the defect frequency is high, our system might be in the $d=3$ dimension, with transition frequency around \SI{100}{GHz}, about \SI{4e-4}{eV}, with crossing temperature $T^*=\SI{5}{K}$. Therefore, for loss channels below \SI{5}{K}, the STM is consistent with the trend that we observed. The second subtle difference is that for quantum TLS, the tunneling rate is lower-bounded by the single-phonon coupling when the temperature is really low; for configurational hopping in the Arrhenius case, since the two levels are near symmetric, the hopping rate follows thermal scaling to much lower temperatures. However this subtle difference is barely noticible in distributed models. Thus, the activation energy that we obtained in the Arrhenius model fitting can also be used for describing quantum TLS defects with energy splitting of $E_a$.  

\subsection{TLS candidates}

Here we discuss the most probable TLS candidates for the observed TLS systems at \SI{e-4}{eV} and \SI{e-5}{eV}. We believe that the \SI{e-4}{eV} defect is most likely from the C-O-C interstitial defect, specifically the O-rotor motion along the symmetry axis. This corresponds to the quantum TLS model with the level splitting determined by the sombrero barrier height~\cite{Thiering16}, which is primarily modulated by normal strain components along the crystal axis. We believe that the \SI{e-5}{eV} defect could come from the neutral spin-1/2 OVH$_3$ complex~\cite{Thiering16}, where the TLS state is defined by the Jahn-Teller distortion-induced modulation of the hyperfine interactions and associated configurational changes~\cite{Kuate21}. Explicitly, based on our DFT calculations, we believe that one $^{17}$O, having diagonal components of the hyperfine tensor with absolute values between 900~MHz and 1000~MHz, and one $^{13}$C atom, having diagonal components of the hyperfine tensor with absolute values between 50 MHz and 200 MHz, within the OVH$_3$ structure generate the TLS. This candidate system has significant insensitivity to magnetic fields, consistent with experimental observations. Intuitively, we believe that when the TLS states show a strong level-asymmetry, the defect will be more sensitive to shear strain; when the initial states are almost degenerate, where level splitting is primarily due to tunneling, then the defect will be sensitive to normal strain. 

We now further expand our argument for the OVH$_3$ as the \SI{e-5}{eV} candidate. Our DFT calculations employed the HSE06 functional for the exchange-correlation energy of electrons with the original parameters~\cite{Heyd2003,Krukau2006}. Convergence was achieved when the magnitude of the Hellmann-Feynman forces was below 10$^{-4}$~eV/\AA~on each atom for a 512-atom ($4\times4\times4$) supercell. We used a cutoff energy of 500~eV and $\Gamma$-point integration. We assume a (100~nm)$^3$ volume and that oxygen infusion happens at every surface carbon atom. Given an 8-atom conventional unit cell with an HSE06 lattice constant of 3.54~\AA, there will be approximately 300000 surface carbon atoms associated with the (100~nm)$^3$ volume. We then have approximately 0.0003 OVH$_3$ defects containing three $^{13}$C and one $^{17}$O; 0.03 OVH$_3$ defects containing two $^{13}$C and one $^{17}$O; 3 OVH$_3$ defects containing one $^{13}$C and one $^{17}$O; 300 OVH$_3$ defects containing one $^{17}$O; 0.3 OVH$_3$ defects containing three $^{13}$C; 30 OVH$_3$ defects containing two $^{13}$C; and 3000 OVH$_3$ defects containing one $^{13}$C where we assume that the natural percentages are 0.1\% for $^{17}$O and 1\% for $^{13}$C. Only the first four isotope proposals lead to the requisite \SI{e-5}{eV} TLS, and of those proposals, only the last two lead to a defect concentration on the order observed in experiment. We have performed calculations ruling out the ground-state splitting of the OVH$_3$ complex as the origin of the \SI{e-5}{eV} defect, as we calculate the contribution of spin-orbit coupling to the ground-state splitting to be bounded below by approximately \SI{e-3}{eV}. 

\begin{figure}[t]
    \includegraphics[width = 0.495\textwidth, page = 1]{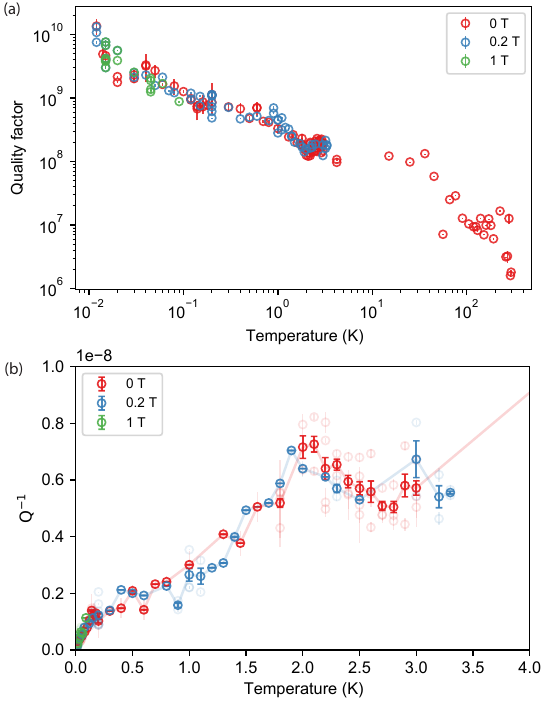}  
    \caption{Temperature-dependence of quality factors of a mm-long segmented beam resonator under different magnetic field amplitudes. \textbf{(a)} Wide-span individual data points, showing no significant magnetic field-dependence up to \SI{1}{T} at temperature $<\SI{4}{K}$. We suspect the high-temperature ($>\SI{100}{K}$) data points to be limited by the gas damping due to the installation of the vector magnet.  \textbf{(b)} Narrow-span averaged data points around \SI{2}{K}, showing no significant magnetic field-dependence on the topological dissipation peak, expected from the properties of \textsuperscript{4}He. }
  \label{fig:SI_BF}
\end{figure}

We rule out the dissipation channels due to coupling to TLS defined by an electronic spin transition, using a vector magnet to probe potential quality factor changes up to \SI{1}{T}, shown in Fig.~\ref{fig:SI_BF}. This does not rule out nuclear spin physics, since the energy scale of a \textsuperscript{13}C nuclear spin under \SI{1}{T} is only \SI{10}{MHz}. 

\section{Superfluid theory and experiment characterization}

The two-fluid model describes liquid $^4\mathrm{He}$ below the superfluid transition temperature as the coexistence of two interpenetrating components: a viscous normal fluid and a superfluid with negligible viscosity. In bulk superfluids, the normal component, with density $\rho_n(T)$, carries all the entropy and behaves like a conventional fluid with finite viscosity, while the superfluid component, with density $\rho_s(T)$, flows without dissipation and exhibits quantum phenomena such as persistent currents and quantized vortices. The total density satisfies $\rho = \rho_n + \rho_s$, and the temperature dependence of $\rho_n$ and $\rho_s$ leads to $\rho_s \rightarrow 0$ near the $\lambda$-point and $\rho_s \rightarrow \rho$ as $T \rightarrow 0$. For a bulk superfluid in or near thermal equilibrium, vortex rings occur only on very small size scales, and their effects can be easily included as a contribution to $\rho_n$ for flows on any macroscopic scale. 

In thin superfluid films, by contrast, thermally-excited vortex-antivortex pairs can occur with somewhat larger separations, and it may be useful to take their motion explicitly into account. In particular, one finds that there exists a topological transition temperature   $T_\mathrm{BKT}$,  lower than the bare transition temperature $T_\mathrm{BKT}$, such that 
the dc flow resistance vanishes only for $T< T_\mathrm{BKT}$.  Dissipation at finite frequencies is non-zero and is dominated by the induced polarization of vortex pairs with appropriate separations. For $T> T_\mathrm{BKT}$, the dc flow resistance is non-zero, and it is dominated by the flow of unbound vortices, for $T<T_\mathrm{BKT}$.

Here, we review the effects of vortex motion on superfluid dynamics in the thin-film limit, and we analyze the effects on resonator dissipation. These effects are special to the context of topological phase transitions and are absent in conventional 3D superfluid transitions. 

\subsection{Normal component dynamics}

We first note that in the normal state, the helium film will follow any motion of the substrate on a time scale very short compared to the oscillation period for the frequencies of interest to us  (order \SI{100}{kHz}). Therefore, there will be negligible hysteresis and no dissipation due to helium flow.

Consider a normal (non-superfluid) helium film of thickness $h\!\sim\!1~\mathrm{nm}$ on a nanobeam oscillating in its fundamental flexural mode. 
The normal component of the liquid helium behaves as a viscous liquid with dynamic viscosity $\eta\approx10^{-6}~\mathrm{Pa\cdot s}$ and volume density $\rho_\mathrm{3D}\approx125~\mathrm{kg/m^3}$.

We want to know the timescale at which the shear viscosity pins down the liquid thin film to the substrate. 
Let the substrate lie in the $xy$-plane and denote by $z$ the coordinate normal to the substrate, with $0 \le z \le h.$ We consider a situation where the substrate oscillates in the $x$-direction, with a velocity $w(t)$ independent of $x$ and $y$.  Then the fluid will have acquired a  velocity field of the form
$\mathbf{v}(x,y,z,t) = u(z,t)\,\hat{\mathbf{x}}$.
The $x$-component of the Navier-Stokes equation then reduces to
\begin{equation*}
\rho_\mathrm{3D} \,\frac{\partial u}{\partial t} = \eta \,\frac{\partial^2 u}{\partial z^2},
\label{eq:tangential-stokes}
\end{equation*}
with the boundary  conditions that $u(0,t) = w(t)$ and that $\partial u/ \partial z = 0$ (no tangential stress) at $z=h$. The result is that shear will relax across the film thickness with a characteristic time $\tau_\perp = h^2 \rho_\mathrm{3D} / \eta \sim\SI{e-12}{s}$, much faster than the oscillatory motion of our resonators. This result will be unchanged if we include the effects of lateral gradients on the length scales of interest for the resonators. Therefore, the normal component of the liquid helium is fully relaxed and tightly pinned down to the substrate by the shear viscosity. 

\subsection{Superfluid dissipation dynamics on a flat substrate}

We begin by considering the properties of a helium film on a flat substrate, which we will allow to undergo small motional oscillations but no distortions. 
Here we follow Ref~\cite{Ambegaokar80}, and we briefly describe the key physics of dynamical BKT systems.  Below the lambda transition, the liquid helium mass density  can be described as
\begin{equation*}
    \rho(\vb r,t) = \rho_s(\vb r,t) + \rho_n(\vb r,t),\qquad \vb j = \rho_s \vb v_s + \rho_n \vb v_n,
\end{equation*}
where $\rho_{s,n}$ and $\vb v_{s,n}$ are superfluid/normal densities and velocities.
We use a frame of reference moving with the substrate, so that $\vb v_s$ and $\vb v_n$ are velocities relative to the substrate and $\vb j$ is the resulting two-dimensional current-density along the surface.  

The superfluid component is irrotational except at singularities and can be written in terms of a phase field $\theta$:
\begin{equation*}
    \vb v_s = \frac{\hbar}{m}\nabla \theta.
\end{equation*}
Single-valuedness of the macroscopic wavefunction $\Psi=\sqrt{n_s}e^{i\theta}$ implies quantized circulation
\begin{equation*}
    \oint \vb v_s\cdot d\vb l = \kappa\, n,\qquad \kappa\equiv \frac{h}{m},\quad n\in\mathbb Z.
\end{equation*}
In a thin film, vortices are point-like objects carrying circulation $\pm \kappa$. For small-amplitude phase variations, density is nearly constant, and the kinetic energy per unit area is
\begin{equation*}
    \mathcal H[\theta]=\frac{J}{2}\int d^2 r\, (\nabla \theta)^2,\qquad
    J\equiv \frac{\hbar^2 \rho_s}{m^2},
    \label{eq:XY-H}
\end{equation*}
where $J$ is the phase stiffness. 
Vortices are topological defects where $\oint \nabla \theta\cdot d\vb l=2\pi q$ with $q=\pm1$. The velocity around a single vortex at $\vb r_i$ (charge $q_i=\pm 1$) is
\begin{equation*}
    \vb v_s(\vb r)=\frac{\hbar}{m}\,\nabla \theta
    = \frac{\hbar}{m}\, q_i \frac{\hat{\vb z}\times(\vb r-\vb r_i)}{|\vb r-\vb r_i|^2}.
\end{equation*}
Substituting $\vb v_s$ into $\mathcal H$ and integrating by parts yields the energy of $N$ vortices (neglecting a core region of radius $a$):
\begin{equation}
    E(\{q_i,\vb r_i\})= 2\pi J\sum_{i<j} q_i q_j \ln\frac{r_{ij}}{a} + N E_c,
    \label{eq:log-interaction}
\end{equation}
with $r_{ij}=|\vb r_i-\vb r_j|$ the vortex separation distance (labeled $d$ in the main text to distinguish from aspect ratio $r$) and $E_c$ the vortex-core energy.
The logarithmic potential is the Green's function of the 2D Laplacian; hence, vortices map to a neutral 2D Coulomb gas.

The classical partition function that sums over all neutral configurations is
\begin{equation*}
    Z = \sum_{N=0}^{\infty}\frac{y^{2N}}{(N!)^2}\int \prod_{i=1}^{2N} d^2 r_i\;
    \exp\!\left[-\beta\, E(\{q_i,\vb r_i\})\right],
    \label{eq:Z-cg}
\end{equation*}
where $y\equiv e^{-\beta E_c}$ is the vortex fugacity.  Neutrality (equal $\pm1$ charges) is required in a finite system.

Because the polarization of small vortex-antivortex pairs will tend to screen the interaction between pairs with larger separation, one finds that at larger length scales,  the effective stiffness $J$ and fugacity $y$ are renormalized. Therefore, we want to integrate out small dipoles with separations between $a e^l$ and $a e^{l+dl}$, i.e., a coarse-grain treatment. With a dimensionless form, $K\equiv J /k_B T$, 
Kosterlitz and Thouless obtained
\begin{align*}
    \frac{d K^{-1}}{dl} &= 4\pi^3 y^2, \\
    \frac{dy}{dl} &= (2-\pi K)\,y, 
\end{align*}
with $l\equiv \ln(r/a)$ the logarithmic length scale. At a critical temperature, where $K = K_c \equiv 2/\pi$, the Kosterlitz-Thouless equations lead to a universal jump in the dc long-wavelength areal superfluid density $\rho_s$, with 
\begin{equation}
    \rho_s(T_{\mathrm{BKT}})=\frac{2 m^2 k_B T_{\mathrm{BKT}}}{\pi \hbar^2},
\end{equation}
just below $T_{\mathrm{BKT}}$, and $\rho_s =0$ above $T_{\mathrm{BKT}}$. 
As one approaches  $T_{\mathrm{BKT}}$ from above, one finds that the correlation length for fluctuations in the superfluid order parameter diverges as
\begin{equation}
    \xi_+(T) = a \exp\!\left(\frac{b}{\sqrt{t'}}\right),\qquad t'\equiv 1-\frac{T_{\mathrm{BKT}}}{T}
    \label{eq:xi-bkt}
\end{equation}
where the precise constant ($a$, $b$) depends on microscopic parameters, and normalized temperature $t'$ needs to be distinguished from time variable $t$.

In the following, we first discuss the diffusion dynamics of an isolated vortex-antivortex pair and then examine how this leads to dynamical damping of superfluid motion. Let $\mathbf{r}_\pm$ be the positions of a vortex and an antivortex, and define their relative coordinate $\mathbf{r}=\mathbf{r}_+ - \mathbf{r}_-$.
Because the vortex core inertia is negligible, the motion is overdamped and governed by a Smoluchowski equation for the probability density $\psi(\mathbf{r},t)$ of the relative separation $\mathbf{r}$:
\begin{equation*}
\partial_t \psi
 = D\,\nabla_{\mathbf{r}}\!\cdot\!\left[\nabla_{\mathbf{r}}\psi
 + \psi\,\nabla_{\mathbf{r}}U(r)\right]
 - \nabla_{\mathbf{r}}\!\cdot\!\big(\mu_{\mathrm{v}}  \,\psi\,\mathbf{F}_{\mathrm{drive}}(t)\big),
\end{equation*}
where $D$ is the diffusion constant of the pair's relative coordinate, $\mu_{\mathrm{v}}  = D/k_B T$  by the Einstein relation, $U(r)=2\pi K\,\ln(r/a)$ is the logarithmic vortex interaction potential, and $\mathbf{F}_{\mathrm{drive}}(t)$ represents the force on the vortices exerted by the oscillatory externally driven superflow.  In the Coulomb charge analogy,  $\mathbf{F}_{\mathrm{drive}}(t)$ is the applied electric field, multiplied by the Coulomb charge.  For the superfluid, $\mathbf{F}_{\mathrm{drive}}(t)$, has a magnitude proportional to the background superfluid velocity, and it is oriented perpendicular to that velocity.  This causes vortices of opposite signs to move in opposite directions perpendicular to the background velocity, which leads to a reduction in the component of $\vb j$ parallel to this velocity. 

In equilibrium, the probability distribution $\psi_0(r)\propto r^{-2\pi K}$ is isotropic. In the following, we assume that the vortex separation $r$ is stabilized by the interaction potential, and only the orientation of the vortex is changing. Note that this is not strictly rigorous, and the result is subjected to a correction factor when considering modulation of $r$ as well. 
Under an AC drive, the first nonvanishing perturbation enters in the $m=1$ angular harmonic, since the drive tends to align the dipole moment $\mathbf{p}=\mathbf{r}$ with the flow:
\begin{equation*}
\psi(r,\theta,t) = \psi_0(r)\big[1 + u(r,t)\cos\theta + \text{higher orders}\big].
\end{equation*}
Here, $\theta$ is the orientation of the vortex pair dipole. So effectively, the drive is reorienting the dipole ensembles. Inserting this heuristic solution into the Fokker--Planck equation and projecting onto the $\cos\theta$ component yields a relaxation equation for the oscillation amplitude~$u(r,t)$,
\begin{equation*}
\partial_t u(r,t) = -\gamma(r)\,u(r,t) + \text{driving}.
\end{equation*}
Dimensional analysis of the relaxation rate shows
\begin{equation*}
\gamma(r) = \frac{c\,D}{r^2},
\end{equation*}
where the numerical constant $c\approx14$ is determined more precisely by the full calculation by Ref~\cite{Ambegaokar80} (AHNS), including the logarithmic interaction and other vortex motions besides the orientational diffusion. The higher order angular harmonics $m$ diffuse much faster since $\nabla_\theta^2 e^{i m \theta} = -m^2 e^{i m \theta}$, and overlap less with the background flow drive. Therefore, the main contribution comes from the $m=1$ first harmonic motion. 

The relaxation time for a pair of size $r$ is thus
\begin{equation}
\tau(r)=\gamma^{-1}(r)\approx \frac{r^2}{14\,D}.
\end{equation}
In an oscillatory experiment at angular frequency $\omega$, the response of such a dipole has the Debye form
\begin{equation*}
\chi(r,\omega)\propto \frac{1}{\gamma(r)-i\omega}
= \frac{\tau(r)}{1-i\omega\tau(r)}.
\end{equation*}
The dissipative/imaginary part of this susceptibility is maximized when $\omega\tau(r)\simeq 1$, which defines a dynamic length scale:
\begin{equation}
L_\omega = \sqrt{\frac{c\,D}{\omega}}
           \simeq \sqrt{\frac{14\,D}{\omega}},
\qquad
l_\omega \equiv \ln\!\frac{L_\omega}{a}
          = \tfrac12\ln\!\frac{14D}{a^2\omega}.
\end{equation}
Vortex pairs with separation $r\lesssim L_\omega$ can follow the AC drive and contribute in phase to the dielectric screening.  Pairs with $r\gg L_\omega$ behave like free vortices, and they can contribute, in principle, to the screening.  However, the number of pairs falls off sharply with increasing $r$ for $T<T_{BKT}$, so they do not contribute much to the overall result. 

In the Coulomb gas analogy,  the linear response to an applied electric field is characterized by a complex frequency-dependent dielectric function, $\varepsilon (\omega, T)$. For the superfluid film, the equivalent relation is
\begin{equation}
\label{jvsep} 
  \mathbf j(\omega)=\frac{\rho_0}{\varepsilon(\omega,T)}\,\mathbf v_s(\omega),
\end{equation}
where $\rho_0$ is the areal superfluid mass density that would have occurred in the absence of the vortex polarizability.  The real part of the factor $\rho_0 / \varepsilon$ may be interpreted as a renormalized superfluid density.

The bound-pair dielectric $\varepsilon_b(\omega)$ is obtained by integrating the response $\big(1-i\omega\tau(r)\big)^{-1}$ over all pair separations $r$ with the weighting factor proportional to the density of such pairs. This leads to
\begin{equation}
\varepsilon_b(\omega,T) - 1
=
\frac{\rho_0}{\rho_s(T_\mathrm{BKT})}\frac{T_\mathrm{BKT}}{T}
\int_{0}^{\infty}\!dl\;{\big[4\pi y(l)\big]^2}\,
\frac{1}{1 - i\,\omega\tau(l)}\label{ebint}
\end{equation}
where $y(l)$ is the vortex fugacity, obtained from the Kosterlitz-Thouless renormalization group equations. 
Evaluating the (static) Coulomb-gas dielectric at the dynamic stop scale $l_\omega$ gives the bound-pair contribution
\begin{align}
    \mathrm{Re}\,\varepsilon_b(\omega,T) &= \frac{\rho_0}{\rho_s(T_\mathrm{BKT})}\frac{T_\mathrm{BKT}}{T}\left[1-\frac{x(l_\omega)}{2}\right], \label{eq:epsbre}\\
    \mathrm{Im}\,\varepsilon_b(\omega,T) &= \frac{\pi}{2}\frac{\rho_0}{\rho_s(T_\mathrm{BKT})}\frac{T_\mathrm{BKT}}{T}\left[4\pi y(l_\omega)\right]^2, \label{eq:epsbim}
\end{align}
where $x(l)\equiv 2-\pi K(l)$. The functions $x(l)$ and $y(l)$ are obtained by integrating from $l=0$ to $l=l_\omega$. 
Near criticality, one can obtain 
closed-form expressions: 
\begin{gather*}
    \text{for } T<T_\mathrm{BKT}:\,\\ 
    x(l)=x_p\coth\!\big(x_p l + \coth^{-1}(x_i/x_p)\big),\, \\
    4\pi y(l)=x_p\mathrm{csch}\!\big(x_p l+ \coth^{-1}(x_i/x_p)\big ),\\
    \text{for } T>T_\mathrm{BKT}:\,\\ 
    x(l)=x_p\cot\!\big(x_p l+ \coth^{-1}(x_i/x_p)\big),\, \\
    4\pi y(l)=x_p\csc\!\big(x_p l+ \coth^{-1}(x_i/x_p)\big),
\end{gather*}
where $x_p=\frac{b}{2}\sqrt{|1- T/T_\mathrm{BKT}|}$, and the constant $x_i=x(l=0)$ depends on the core energy and the precise choice of the starting radius $a$. 
The parameter $b$ is the same as in Eq. (\ref{eq:xi-bkt}).

Above $T_\mathrm{BKT}$, when $L_\omega\gtrsim\xi_+(T)$, i.e., the participating vortex separation is beyond the correlation length $\xi_+(T)$, the pairs are no longer physically bound, and the dissipation is dominated by the free-vortex channel, which has the form~\cite{Bishop80}
\begin{equation}
   \varepsilon_f=\frac{4\pi^2 iD\hbar^2\rho_0}{\omega m^2 k_BT}n_f 
 \end{equation} 
where $n_f \approx \xi_{+}^{-2}$ is the density of free vortices. 
Beyond this threshold, we replace the bound dielectric response with this free-vortex response function.

So far, we have considered the behavior of a film on an infinite planar surface.  In our application, however,  the substrate is a beam that is very long in one direction but has a finite width and thickness of order several hundred nm.  At sufficiently low frequencies, where the length $L_\omega$ exceeds the width $W$, the two-dimensional description will break down, and a quasi-one-dimensional analysis must be used. We shall return to this issue later, but for the moment, we shall assume the frequency is high enough that we can use the dielectric function obtained from the two-dimensional description.

\subsection{Vortex-induced dissipation of resonators}

Our expressions for the dielectric response function $\varepsilon(\omega, T)$  can be used to calculate the dissipation effect of the vortex dynamics on the resonator motions. However, we must now take into account that in our geometry, the tangential velocity of the substrate depends on position; it cannot be eliminated by transforming to a uniform moving frame.  Instead, it is convenient to write
\begin{equation}
\vb v_s (\vb r) = [ \vb u_s (\vb r)  - \vb v_m (\vb r) ]_{\parallel},\label{eq:rel_v}
\end{equation}     
where  $\vb u_s (\vb r)$ is the superfluid velocity in the laboratory frame, $\vb  v_m (\vb r)$ is the local velocity of the substrate (and also of the normal fluid ), also in the laboratory frame, and $[...]_{\parallel}$ indicates the vector component parallel to the plane of the surface. 

The quantity $ [ \vb u_s ]_{\parallel}$  can be written as 
\begin{equation*}
[\vb u_s ] _{\parallel} =  (\hbar / m)   \nabla \phi (\vb r) ,
\end{equation*}
where $\phi$ is the phase of the (locally averaged) superfluid order parameter. In turn, the time-dependence of $\phi$ is governed by the Josephson equation, 
\begin{equation}
 \hbar \frac {\partial } {\partial t} \phi (\vb r, t) = \mu (\vb r, t) ,\label{eq:JE}
\end{equation}
where $\mu$ is the local chemical potential of the helium film. This includes an external portion $\mu_\mathrm{ext}$ arising from the attraction to the substrate, and an internal portion $\mu_\mathrm{int}$  depending on the local density and temperature of the helium. In the regime of interest to us, we can consider $\mu$ to be constant across the thickness $h$ of the film, though its constituent parts will vary. However, we must take into account the variation of $\mu$ along the surface of the substrate.    

At sufficiently low temperatures, where exchange with the vapor is negligible, for a locally flat substrate, the value of $\mu_\mathrm{int}$ at the free surface of the film should be independent of position, and essentially the same as for a bulk helium bath. The external piece, due to the interaction with the substrate, evaluated at the free surface, is related to the local film thickness by  

\begin{equation}
\frac{\mu_\mathrm{ext}} {v_0}  =  - \frac {A_H} { 6 \pi h^3} ,
\end{equation} 
where $A_H$ is the Hamaker constant and $v_0$ is the volume occupied by a helium atom. More precisely, one should define $v_0 = m \,dh/d\rho$, where $\rho$ is the areal helium density. We shall set $A_H \approx 2 \times 10^{-20} $J, borrowed from Si-He interface results. On a curved surface, there can be curvature corrections to the Hamaker constant as well as corrections to $\mu_\mathrm{int}$ arising from the surface tension of helium, but we shall ignore them here.

On a substrate with non-uniform motion, it remains true that the local superfluid current relative to the substrate, at frequency $\omega$, is given by Eq.~(\ref{jvsep}). Hence, we can calculate the dissipated power per unit area as
\begin{equation*}
  P= \big\langle\mathbf j\cdot \dot{\mathbf v}_s\big\rangle_t=\frac{\omega\rho_0}{2}\,|\mathbf v_s|^2\,\mathrm{Im}[-\varepsilon(\omega,T)^{-1}].
\end{equation*}
Therefore, the vortex-induced dissipation takes the form
\begin{gather}
  Q^{-1} = \frac{\rho_\mathrm{0}\iint dS|\mathbf v_s|^2}{\rho_\mathrm{dia}\iiint dV|\mathbf v_m|^2}\mathrm{Im}[-\varepsilon(\omega,T)^{-1}]\nonumber\\
  =\frac{m_s}{m_\mathrm{eff}}\frac{\langle|\mathbf v_s|^2|\rangle}{v_{m,\mathrm{max}}^2}\mathrm{Im}[-\varepsilon(\omega,T)^{-1}],\label{eq:Q}
\end{gather}
with an equivalent frequency response
\begin{gather}
  \frac{\Delta f_m}{f_m} = \frac{m_s}{2m_\mathrm{eff}}\frac{\langle|\mathbf v_s|^2|\rangle}{v_{m,\mathrm{max}}^2}\mathrm{Re}[\varepsilon(\omega,T)^{-1}],
\end{gather}
where $m_s$ is the superfluid mass and $m_\mathrm{eff}$ the effective mass of the resonator mode. 

Further calculation requires the evaluation of the effective velocity ratio ${\langle|\mathbf v_s|^2|\rangle}/{v_{m,\mathrm{max}}^2}$ between the average superflow velocity power ${\langle|\mathbf v_s|^2|\rangle}$ on the resonator surface in response to a resonator velocity power ${v_{m,\mathrm{max}}^2}$. Thus, we need to calculate the superfluid velocity field on the surface of the resonator based on the oscillator oscillation profile. In the following, we derive the equations that govern the superflow of the film driven by the inertial force from the resonator motions.

We consider a superfluid He film of thickness $h(\mathbf s,t)$ coating a solid body whose surface is $S$. The surface coordinate is $\mathbf s = (s_1,s_2)$, with associated surface gradient $\nabla_S$ and surface Laplacian $\Delta_S = \nabla_S^2$. The film has mass density $\rho_0$ and tangential superfluid velocity $\mathbf v_s(\mathbf s,t)$.

The body undergoes a general (time-dependent) motion that produces an effective tangential acceleration field 
$\mathbf a_{\rm ext,\parallel}(\mathbf s,t) = \partial \, \vb v_m (s,t)  / \partial t$ 
in the lab frame, evaluated at the surface and projected onto the surface tangent plane. This field can come from translations, rotations, and their combinations; it need not be the gradient of any scalar potential.

The linearized  continuity equation on $S$, for small amplitude perturbations, is 
\begin{equation}
  - i \omega h +  \frac {h' \rho_0 }{\varepsilon (\omega)} \nabla_S\cdot \mathbf v_s  = 0,
  \label{eq:cont-general}
\end{equation}
where $h' \equiv dh/d\rho$ is the derivative of the film thickness with respect to variations in the areal helium mass density.

Combining Eqs.~(\ref{eq:rel_v}-\ref{eq:JE}), we see that 
\begin{equation} 
i \omega \vb v_s = \frac{1}{m} \nabla_S \mu + \vb a _{\mathrm{ext} , \parallel} \label{eq:momentum}
\end{equation}
Furthermore, we have 
\begin{equation*}
\nabla_S \mu = \nabla_S \mu_\mathrm{ext} (h) = \mu' \nabla_S h,
\end{equation*} 
where $\mu' = d\mu_\mathrm{ext}/dh$, and $h_0$ is the equilibrium film thickness.
We remark that in the limit of low temperatures, we may write $\nabla_S \mu= - (m/\rho_\mathrm{3D} ) \nabla_S p$, where $p$ can be interpreted as a surface pressure.

Typically, normal stress balance at the free surface relates $p$ to the local thickness via a vdW disjoining pressure $\Pi(h)=A_H/(6\pi h^3)$, where $A_H\sim\SI{2e-20}{J}$ is the Hamaker constant (borrowed from Si-He interface results), and the surface tension $\gamma\sim\SI{3.7e-4}{N/m}$,
\begin{equation*}
  p = p_0 + \Pi(h) - \gamma\,\Delta_S h.
\end{equation*}
However, we observe in the simulation that the surface tension term is negligible compared to that of the vdW disjoining pressure. Therefore, we do not include the surface tension term in the following discussion. 

Combining the above equations, we arrive at
\begin{equation*}
\omega^2 \vb v _s = - \omega^2 [\vb v_m]_\parallel + \frac {\rho_0 \mu' h'} { m\varepsilon (\omega) } \nabla_S (\nabla_S \cdot \vb v_s) .
\end{equation*} 

Ignoring the imaginary part of $\varepsilon$, we may define a third sound speed $c_3 = ( \rho_0 h'\mu' /m\varepsilon)^{1/2}\approx({h_0\,\Pi'(h_0)}/{\rho_\mathrm{3D}})^{1/2}$, and we may rewrite this equation as
\begin{equation}
\omega^2 \vb v_s + c_3^2 \nabla_S( \nabla_S \cdot \vb v_s)  =  - \omega^2 [\vb v_m]_\parallel .
\end{equation}

\begin{figure}[t]
    \includegraphics[width = 0.495\textwidth, page = 1]{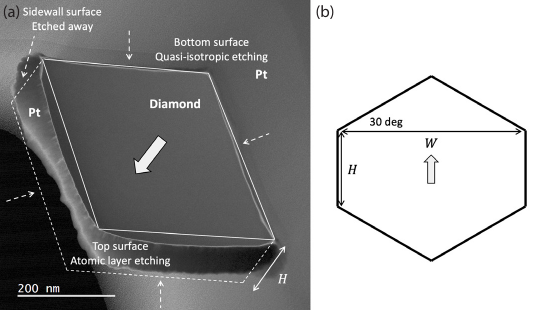}  
    \caption{\textbf{(a)} Transmission electron microscopy (TEM) image of the cross-section (rotated) of a diamond nanobeam embedded in a platinum cladding. The large solid arrow points towards the top surface of the device. The angled bottom and top surfaces are a result of the quasi-isotropic etching (undercut, bottom) and atomic layer etching (ALE, top) over an extended period of time. The dashed lines outline a typical shape of the resonators used in the experiment, where the vertical sidewall is etched away by the ALE process and not visible in this TEM image. \textbf{(b)} The cross-section schematic of a diamond device used for the COMSOL simulation of the superfluid experiment. Here $H = \SI{200}{nm}$, $W=\SI{500}{nm}$. }
  \label{fig:SI_TEM}
\end{figure}

{\subsubsection* {COMSOL simulation}}

These equations above were solved, assuming $\varepsilon (\omega) = 1$, in COMSOL Multiphysics PDE module in the frequency domain at $\omega$, where the inertial acceleration field in the frequency domain is obtained from velocity projection to the tangential surface (normal vector $\hat{\mathbf n}$) through $\mathbf a_{\rm ext,\parallel}(\omega) = \mathbf a_{\rm ext}(\omega)-(\mathbf a_{\rm ext}(\omega)\cdot \hat{\mathbf n})\hat{\mathbf n}$, where $\mathbf a_{\rm ext}(\omega) = -i\omega \mathbf v_{m}(\omega)$. 
In the simulation, the geometry of the resonator is a single beam, pinned at the ends, which has a modified rectangular cross-section with top and bottom surfaces exhibiting a \SI{30}{\deg}-angle wedge shape, with exact dimensions shown in Fig.~\ref{fig:SI_TEM}. This is the result of quasi-isotropic etching and atomic layer etching of the top and bottom surfaces, with the angle retrieved from a cross-section electron microscopy image. 

\begin{figure}[t]
    \includegraphics[width = 0.495\textwidth, page = 1]{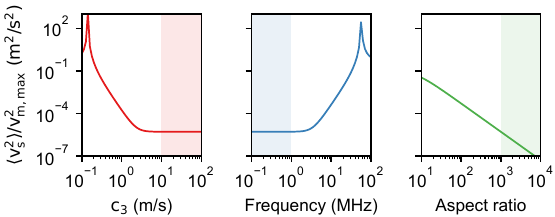}  
    \caption{Simulated third sound velocity, frequency, and aspect ratio scaling, for the relative superfluid kinetic energy. Shaded regions are parameter regimes of interest for this work. }
  \label{fig:SI_SF}
\end{figure}

The resulting superflow velocity field for a fundamental flexural mode is shown in Fig.4 of the main text, while the surface-averaged velocity power ratio ${\langle|\mathbf v_s|^2|\rangle}/{v_{m,\mathrm{max}}^2}\sim \SI{e-6}{}$ required to calculate vortex-induced dissipation is shown in Fig.~\ref{fig:SI_SF}, showing its dependence to key parameters such as $c_\mathrm{3}$, $\omega$, and device aspect ratio. Note that the numerical simulation reveals a ${\langle|\mathbf v_s|^2|\rangle}/{v_{m,\mathrm{max}}^2}=\SI{8.2e-7}{}(500[$\SI{}{\micro\meter}$]/L)^2( {H}/200[$\SI{}{nm}$])^2$ scaling for our device cross-section, where $L$ is the length of the beam and $H$ is the thickness of the beam. This choice of a simplified one-dimensional structure is to provide orders of magnitude estimation of the dissipation effect, since mechanical simulation of the exact structure of a large-scale self-tensioning network is numerically hard to simulate. Note that our system works at the far off-resonant regime from the first resonant mode of the superfluid motion. Here, the surface integral is extended slightly beyond the device structure, as the superfluid is not confined. However, we did find that for high aspect ratio structures, those contributions can be safely neglected. Further, we found that higher-order nonlinear effects such as stationary fluid migration towards oscillator nodes can be safely ignored when the vibration amplitude $A<\SI{1}{\micro\meter}$, with velocity power contribution ${\langle|\mathbf v_s(2\omega)|^2|\rangle}/{v_{m,\mathrm{max}}^2}<\SI{e-6}{}\propto (A/L)^2$. 

\bigskip 

{\subsubsection* {Fitting of the dissipation peaks}}

Combined with the mass ratio and with an assumption for the form of Im $\varepsilon (\omega, T)$, we can use the above results to calculate the dissipation rate, at least under conditions where  $|\varepsilon - 1|$ is small. 
Assuming that the two-dimensional form of $\varepsilon(\omega,T)$ derived earlier can be applied to our system, we fit the expression using Eq.~(\ref{eq:Q}) to the data using $l_\omega$, $T_\mathrm{BKT}$, ${\langle|\mathbf v_s|^2|\rangle}/{v_{m,\mathrm{max}}^2}$, $b$, and $x_i$ as free parameters, independent for all three curves in the paper. 
We obtain a $T_\mathrm{BKT}\approx\SI{1.2}{K}$ for all three modes within 10\% deviations. The velocity power ratios are \SI{1.1e-5}{}, \SI{1.1e-5}{}, \SI{1.0e-5}{}, the $l_\omega$ are \SI{3.5}{}, \SI{3.2}{}, \SI{3.0}{}, the $b$ are \SI{0.50}{}, \SI{0.55}{}, \SI{0.60}{}, and the $x_i$ are \SI{0.89}{}, \SI{0.86}{}, \SI{0.96}{}, for the \SI{137}{kHz}, \SI{187}{kHz}, \SI{267}{kHz} modes respectively. We find that the fit to $x_i$ is not sensitive and very much depends on the initial condition, given that we are in the linearized regime where $l_\omega>1/x_i$.

In the following, we focus on the analysis of the fundamental mode, since we have the best knowledge of its displacement profile. The fitting yields a dimensionless cutoff length $l_\omega = 3.5$ and a velocity power ratio ${\langle|\mathbf v_s|^2|\rangle}/{v_{m,\mathrm{max}}^2}\approx \SI{1.1e-5}{}$, which is about an order of magnitude larger than the numerical simulation result given the device aspect ratio. 
For further analysis, we assume a core size of $a=\SI{1}{nm}$, due to a functional degeneracy between $a$ and $D$, yielding a fitted vortex diffusion constant at $D \approx \SI{3e-11}{m^2/s}$, which leads to the damping effect $Q^{-1}\sim\SI{e-8}{}$ measured in Fig.4 of the main text. Note that this value of diffusivity is two orders of magnitude lower than the one reported in Ref~\cite{Bishop80}. 
In the following, we briefly explain where we believe this discrepancy comes from.

For a finite-size system, the cut-off $L_\omega$ can be much earlier as the vortex separation is constrained by the surface geometry~\cite{Kotsubo86}, which exhibits an effective periodic condition along the circumference of the beam. We derive in a later subsection, \textit{Finite-width impact on vortex motion}, the influence of our device's specific surface geometry as opposed to that of an infinite two-dimensional plane. We find that, in the case of a modified superfluid response from Eq.~(\ref{eqf}), the core size is no longer functionally degenerate with $D$ and can be extracted to be $a\approx\SI{3}{nm}$. The diffusivity instead can be extracted from the amplitude of the dissipation, combined with the results of the simulated velocity power fraction, to be $D\approx\SI{e-8}{m^2/s}$, similar to the literature value reported in Ref~\cite{Bishop80}. Therefore, we believe our system is in the finite-size limit, qualitatively different from those previous studies where the infinite two-dimensional case can be applied. We find that this is a unique feature that only exists for topological dissipation by nanoscale structures.

Apart from the finite sizing effect, we found that the nonlinear damping due to a large vibration amplitude $A\sim\SI{1.8}{\micro\meter}$ could potentially explain the discrepancy when comparing with the two-dimensional model. However, we don't believe this is the case in our experiment situation. Further, an average cross-section asymmetry of the self-tensioning tether on the order of $5\%$ could also account for this discrepancy, by introducing superfluid-lubricated slipping at the adhesion interface, causing relative superflow against the substrate without the need for inertial forces. We further note that other types of motion, such as axial rotation of the beam, could lead to different superfluid flow patterns and, therefore, modification of the velocity power ratio. Here, we neglect their analysis due to the small kinetic energy participation during our oscillator motions. However, it is hard to gain more knowledge on the microscopic details of the superfluid and the mechanical structural dimensions without destructive measures; we are satisfied with the order of magnitude confirmation by the finite-size effect-corrected model. 

\subsubsection*{Other possible mechanisms}

We note that vortex dynamics is not the only possible source of dissipation in superfluid helium films. There have been reports and theories on evaporation-induced dissipation and heat transport-induced dissipation~\cite{Bergman71},  which do not involve vortex physics. However, we find that these effects would be quite negligible in our experiments.

Evaporation-induced dissipation was found to scale as $Q^{-1}\propto h^{11/2}$, and thus negligible at a few atomic layer thickness. For thin films, heat transport leads to more prominent dissipation in the superfluid phase. In the two-fluid model, the superfluid transports without carrying entropy, therefore dynamically modulating local temperature when there is superfluid transport. Such a dynamical temperature gradient leads to thermal diffusion and thus dissipation, leading to a more prominent effect as compared to the intrinsic thermal elastic damping of diamond. Temperature fluctuations are proportional to fractional changes in the local film thickness $h$ rather than fluctuations in $\vb v_s$, so in the electrical analogy, their dissipative effects  are  characterized by  an imaginary contribution to a magnetic permitivity $\mu_m (\omega)$ as opposed to $\varepsilon(\omega)$ 

Based on the theory prediction for a few monolayer-thick, \SI{100}{kHz}-frequency third sound wave on an infinite substrate bulk, the permitivity response is $\mathrm{Im} {[\mu_m]} = \frac{TA_H}{2\sqrt{2}\pi\tilde{L}^2\rho^2h^4}\sqrt{\frac{\rho_\mathrm{sub} C_\mathrm{sub}\kappa_\mathrm{sub}}{\omega}}\propto T^4$, where $\tilde{L}$ is the latent heat of the film, and $C_\mathrm{sub}\,,\kappa_\mathrm{sub}$ are the specific heat and heat conductance of the substrate.  

If the imaginary part of $\varepsilon$ had a contribution of this magnitude, it could lead to excess dissipation on the order of 
$Q^{-1}_\mathrm{th} \sim 10^{-9} (T/1[K])^4$, with $T$ measures in kelvin. In our geometry, however, the energy arising from fluctuations in $h$ is similar in magnitude to the kinetic energy associated with the component of $\vb v_s$ perpendicular to the beam, which is smaller by a factor of order $(W/L)^2 \approx 10^{-6} $ than the kinetic energy associated with flow parallel to the beam. Therefore, the contribution of heat transport to $Q^{-1}$ is predicted to be of order $< 10^{-14}$ in our experiments, which can be safely neglected.

\bigskip
\subsubsection*{Rectangular beam}

We can gain further insight into the physics of the helium flow by examining a simplified geometry, where analytic results are straightforward to obtain.
We consider a finite-wavelength vibrational mode of a beam of infinite length with a rectangular cross-section, of thickness $H$ and width $W$.
We use coordinates where $-W/2 < y < W/2$ and $-H/2 < z < H/2$, and the beam axis lies in the x-direction.
The beam is subjected to a small oscillating displacement of the form $  \vb U_m = \hat{z} U e^{i kx - i\omega t}$, where $\hat {z }$ is a unit vector in the z-direction.

On the top surface where $z=H/2$, one finds 
\begin{equation}
\phi (x, y,H/2)   = \phi_0 e^{ikx -i\omega t} \cos qy ,
\end{equation}
where $q^2 = (\omega / c_3)^2 - k^2$,
$\phi_0 $ is an amplitude to be determined, and we have assumed $ \omega > |c_3 k| $. On the side surface where $y = W/2$, we have 
\begin{equation}
\phi (x, W/2, z)   = \phi_1 e^{ikx -i\omega t} \sin qz  .
\end{equation}
At the edge $(y,z) =  (W/2,H/2)$,  continuity requires
\begin{equation*}
\phi_0 \cos (qW/2) = \phi_1 \sin (qH/2) ,
\end{equation*} 
while conservation of superfluid flow requires
\begin{equation*}
\frac {\hbar}{m} \phi_0 q \sin (q W/2) = \frac {\hbar}{m} \phi_1 q \cos (q W/2) - i \omega U .
\end{equation*} 
These equations lead to the result
\begin{equation}
\label {phi0rect}
\frac {\hbar}{m} \phi_0  = \frac {i \omega U} {q} \left( \frac { \sin(qH/2)} {\cos [q (W+H)/2]} \right).
\end{equation} 

Superflow velocities on the various surfaces can be obtained from the spatial derivatives of $\phi$.  We remark that $\vb v_s$ has opposite directions on the top and bottom surfaces, and in the limit $q (W+H) \ll 1$, this will reduce to
\begin{equation}
\vb v_{s,\mathrm{top}}  = \mp \hat{x} \, kH \, \omega  U /2 .
\end{equation}
We also remark that the denominator of Eq.~(\ref{phi0rect}) will vanish, and the linear response will diverge, when  $q (W+H) = \pi$.  This will occur if $\omega$ reaches a resonant frequency $\omega_r$,  of which the lowest is equal to $\pi c_3 / (W+H) $, in the limit $c_3 k \ll \omega$.

Though this is a traveling-wave computation, the results can also be applied to a beam of finite length $L$, with the appropriate boundary conditions. Working only to linear order in the vibration amplitude, the superfluid response is linear in the drive, so the superfluid velocity field for a standing-wave mode is obtained by superposing the solutions for two counter-propagating traveling waves. If $v_{m,\mathrm{max}}$ denotes the peak mechanical velocity at an antinode, then the spatial average along the beam introduces $\langle \cos^2(kx)\rangle = 1/2$, so that the ratio $\langle |\vb v_s|^2\rangle / v_{m,\mathrm{max}}^2$ is reduced by a factor of $1/2$ relative to the traveling-wave case.

If the beam is pinned at the ends,  we must choose $k = \pi n/L$, where $n$ is a positive integer.   Depending on the boundary conditions for helium flow at the ends of the beam, we may find corrections to the flow near the ends.  However, the added flow fields will fall off exponentially from the ends with a decay length of order $W+H$, so the end effects will be negligible for a wire of sufficient length.

The above equations were also solved, assuming $\varepsilon (\omega) = 1$, in COMSOL Multiphysics PDE module in the frequency domain at $\omega$, where the inertial acceleration field in the frequency domain is obtained from velocity projection to the tangential surface (normal vector $\hat{\mathbf n}$) through $\mathbf a_{\rm ext,\parallel}(\omega) = \mathbf a_{\rm ext}(\omega)-(\mathbf a_{\rm ext}(\omega)\cdot \hat{\mathbf n})\hat{\mathbf n}$, where $\mathbf a_{\rm ext}(\omega) = -i\omega \mathbf v_{m}(\omega)$.  As expected, the numerical results agree with the analytic formulas.

\subsubsection* {Cylindrical wire}
 
We consider here another simple case, the vibrations of a stretched cylinder of radius $R$ and infinite length. We take the wire to lie along the x-axis, subject to a small oscillating displacement of the form $ \vb U_m = \hat{z} U e^{i kx - \omega t}$, where $\hat {z }$ is a unit vector in the z-direction, and $U$ is the amplitude of the oscillation. 
 
If we define the azimuthal angle $\theta$ such that points on the surface of the cylinder have coordinates $z= R \cos \theta$ and $y=R \sin \theta $, then the component of displacement tangent to the surface has the form $\hat{\theta} U_\theta e^{ikx-i \omega t}$, where $ \hat {\theta}$ is a unit vector in the direction of increasing $\theta$ and 
 \begin{equation*}
 U_\theta = - U \sin  \theta 
 \end{equation*} 
 
 Because the unperturbed system has translational symmetry in the x-direction and rotational symmetry about the x-axis, the linear response of the system, characterized by small perturbations in superfluid thickness $\delta h$ and phase $\phi$, will have the form
  \begin{equation}
  \delta h = \eta \cos \theta e^{ikx- i \omega t} , \,\,\, \phi = \Phi \cos \theta e^{ikx - i \omega t} .
  \end{equation}
  These forms will satisfy the equations of motion Eq.~(\ref{eq:rel_v},\ref{eq:cont-general},\ref{eq:momentum}) derived above, provided
  \begin{equation*}
  - i \omega \eta = \frac { h' \rho_0 (k^2 + R^{-2} )} {\varepsilon (\omega)}  \left( \frac {\hbar \Phi}{m} \right) - i \omega U \frac {h' \rho_0} {R \varepsilon (\omega) } ,
  \end{equation*} 
  \begin{equation*}
  i \omega \left( \frac { \hbar  \Phi} {m} \right) = \mu' \, \eta .
  \end{equation*} 
  The solution is 
  \begin{equation*} 
  \frac {\hbar \Phi}{m} =   - i \omega U \frac {c_3^2} { R }  \left( \frac{1}{\omega^2 - \omega_r^2} \right) ,
  \end{equation*}
  \begin{equation*}
  \omega_r^2 = c_3^2 ( k^2 + R^{-2} ) 
  \end{equation*} 
  where $c_3^2=h'\mu'\rho_0/m\varepsilon(\omega)$. 
  The indued superfluid velocity will have components parallel to $\hat{x}$ and $\hat {\theta}$, given by
  \begin{gather} 
  v_{s,x} = - \omega U \cos \theta \frac{ k R } {1 + k^2 R^2} \frac{ 1 } {1 - \frac{\omega^2}{\omega_r^2}} \nonumber\\
  \approx - k R \,\omega U \cos \theta 
  \end{gather}
  \begin{gather}
  v_{s,\theta} = i \omega U \left[ 1  - \frac {\omega_r^2} { (\omega_r^2 - \omega^2) (1 + k^2 R^2) } \right]  \sin \theta \nonumber\\
  \approx\left(1-\frac{c_m^2}{c_3^2}\right)(kR)^2 i \omega U  \sin \theta
  \end{gather} 
  Note that if $kR$ and $\omega/\omega_r$ ar both small compared to unity, we have $v_{s,x}   \sim kR|\vb v_m|$ and $v_{s,\theta} \sim (1-c_m^2/c_3^2)(kR)^2|\vb v_m|$, where $c_m$ is the flexural mode velocity. Therefore, ${\langle|\mathbf v_s|^2|\rangle}/{v_{m,\mathrm{max}}^2}\approx (kR)^2 = \SI{3e-6}{}$ if we choose wavelength $\lambda=\SI{400}{\micro\meter}$ and radius $R=\SI{100}{nm}$. The results are qualitatively consistent with the COMSOL simulation.

\subsubsection*{Finite-width impact on vortex motion}

As mentioned above, our discussion of vortex dynamics based on a two-dimensional geometry will break down at frequencies so low that  $L_\omega$ is comparable to the width of the suspended diamond structure.  We believe this situation is applicable to the current experiments, and it will be relevant for experiments with even lower frequencies, thinner wires, or larger values of the vortex diffusion constant.

For simplicity, let us again consider a cylindrical wire of radius $R$ and infinite length. We can map the surface onto a flat strip of width $W=2\pi R$, with periodic boundary conditions joining the two sides of the strip. This means that a charge at point $(x,y)$ will be accompanied by an infinite set of image charges of the same sign, located at points $(x, y+nW)$, where $n$ is any positive or negative integer. Then, if a pair of unit positive and negative charges has a separation $(X, Y)$, with $X>W$, the force between them will essentially be that between a pair of line charges in two dimensions, which will be independent of $X$. Furthermore, the dependence on $Y$ will fall off exponentially with $X/R$. Thus, we may approximate the interaction energy, for $|X|>R$, as    
\begin{equation}
E(X,Y) \approx \frac {2 \pi K_R T }{ R} (|X| - R)  + F_R ,
\end{equation}
where $K_R T$  is the renormalized point-charge coupling, taking into account screening by vortex pairs with separation less than $R$, and $F_R$ is the free-energy cost to create a pair with separation $R$. We may identify $K_R$ with the value of $K(l) $ obtained from the two-dimensional Kosterlitz-Thouless equations at length scale $L=R$.

Since $E$ is independent of $Y$, the vortex pair will respond to a uniform electric field in the $y$-direction like a pair of free charges, for as long as the pair continues to have X-separation greater than $R$.   The total density of these quasi-free vortex pairs may be estimated as 
\begin{equation*}
n_{\rm{qfp}} \approx 2 a^{-4}  \int_0^{2 \pi R} dY \int_R^\infty dX e^ {- E(X,Y) / k_B T }  
\end{equation*} 
\begin{equation}
= \frac{2 R^2  a^{-4}}{K_R}  e^{-F_R / k_B T } .
\end{equation}
Further, we may estimate $F_R$  with
\begin{equation}
e^{-F_R / k_B T} \approx ( 4 \pi y_R)^2 (a/R)^4 ,
\end{equation} 
where $y_R$ is the value of $y(l)$ that one obtains from the two-dimensional equations at length scale $L = R$. Now, taking into account that the number of quasi-free vortices should be twice the number of quasi-free pairs, we may estimate the contribution of the quasi-free vortices to the dielectric constant at frequency $\omega$, for an electric field perpendicular to $\hat {x}$, as 
\begin{equation}
\label{eqf} 
{\rm{Im}} \, \varepsilon_{\rm{qfp}}  \approx \frac {8\pi^2 D \hbar ^2 \rho_0 n_{\rm{qfp} }} {\omega m^2 k_b T }  \approx  \, C \frac { \omega_R} {\omega} \rm{Im}\, \varepsilon_b (\omega_R,T), 
\end{equation} 
where $\omega_R \equiv 14D / R^2$. In the last expression, we note that the prefactor $C=1$ is chosen such that the function follows continuously to the limit of the two-dimensional case when $\omega>\omega_R$. 

We may also estimate the contribution to Im$[\varepsilon (\omega)]$ from bound pairs with separation less than $R$ by truncating the integral in Eq.~(\ref{ebint}) 
at $l = \ln (R/a) $.   This gives a result which is smaller than $\epsilon_{\rm{qfp}} $ if $\omega$ is small compared to $\omega_R$, so it may be ignored in this regime. 
The net result is that we should use the two-dimensional result Eq.~(\ref{eq:epsbim}) for Im$[\varepsilon(\omega)]$ if   $\omega \gg \omega_R$, but we should use Eq.~(\ref{eqf}) if 
$\omega \ll \omega_R$. However, there may be a large crossover regime that needs further study.

When applying these results to superfluid flow, we must recall that the superfluid velocity $\vb v_s$ is perpendicular to the analog electric field $\vb E$. The results derived above are therefore applicable to $\vb v_s$ in the x-direction, parallel to the axis of the wire. This is the situation applicable in our experiments, where the largest superfluid velocities are in the direction parallel to the wire.

The case where there is a uniform electric field in the x-direction is also of possible interest. This corresponds to uniform $\vb v_s$ in the y-direction. A situation that would arise on the surface of a cylinder with an oscillating rotation about its axis. In this case, the induced vortex current will be limited by the energy cost to separate vortex pairs along the x-axis. We estimate that dissipation at low frequencies will be reduced by a factor of order $(\omega/ \omega_R )^2$ in this case.

The analyses of this subsection should also apply to a wire whose cross-section is non-circular, if we replace $R$ by $C/2 \pi$, where $C$ is the perimeter of the cross-section. Applying this result to the cut-off length retrieved from our data (logarithmic $l=3.5$), we found an effective vortex core size of $a \approx \SI{3}{nm}$. Assuming the flow is dominated by the longitudial direction, the analytical velocity power fraction is ${\langle|\mathbf v_s|^2|\rangle}/{v_{m,\mathrm{max}}^2}\approx (kH/2)^2= \SI{e-6}{}$ with device thickness $H = \SI{200}{nm}$ and a characteristic length of $L=\SI{400}{\micro\meter}$. Using the fitted dissipation amplitude from the fundamental mode data, we extract a $\omega_R/2\pi \approx \SI{1.5}{MHz}\gg\omega/2\pi=\SI{137}{kHz}$, and a vortex diffusivity $D = \omega_R C^2/56\pi^2\approx\SI{e-8}{m^2/s}$, similar to the following literature value~\cite{Bishop80}. We point out that the value $D$ here is close to the value $\hbar/m_\mathrm{He}=\SI{1.6e-8}{m^2/s}$, which is the scale of the maximum microscopic vortex diffusion process of quantum mechanical nature.

\subsection{Goldstone–mode–induced thinning of the superfluid film}

Here, we review the results of the theory of thinning proposed in Ref. \cite{Zandi04}. The film is assumed to be in thermal equilibrium with a reservoir at a chemical potential $\mu$. The film thickness $h$ is then found to obey the equation
\begin{equation}
\frac{A_H}{6\pi h^3}-\Theta(T)   \frac{k_BT}{h^3}= -  \frac{\mu_{\mathrm{eff}} }{v_0}  ,
\end{equation}
where $\mu_{\mathrm{eff}}$ is the difference between $\mu$ and the chemical potential to keep a reservoir of bulk helium in equilibrium with its vapor at the given temperature, $v_0$ is the volume per helium atom, and $\Theta(T>T_\mathrm{BKT})=0$ represents the vanishing acoustic Casimir effect from the normal fluid phase, while $\Theta(T<T_\mathrm{BKT})=\Theta_0$ represents the finite Casimir effect due to the emergence of the superfluid phase. 
The analysis of Ref. \cite{Zandi04} considers both bulk fluctuations and surface ({\it{i.e}}, third sound) modes, leading to a predicted value $\Theta_0 \approx 0.15$ (while experimentally measured $0.3$). By assuming a constant chemical potential across the transition temperature, this leads to an effective fractional thickness reduction of 
\begin{equation}
    \frac{\delta h}{h}=\frac{2\pi\Theta_0 k_BT}{A_H}.
\end{equation}
Here, the constant chemical potential represents the competition for a finite amount of He atoms between surfaces of different materials and sizes; therefore, the change is due to the difference in potential between the film on diamond and the film on the reservoir surface, with the bulk contribution perfectly cancelled. We assume the effective reservoir has a much larger surface area than the diamond surface, and a weaker Casimir effect, and thereby maintains the thickness across the transition after re-equilibrium. In a more realistic setting, the $A_H$ and $\Theta_0$ parameters can be modified by the reservoir behavior, but the functional form remains the same. 

Compared to literature works~\cite{Zandi04} with thickness at \SI{20}{nm}, our system is significantly thinner in film thickness, therefore, it could require additional modification of the presented theory. 
Note that closer to the transition temperature, due to the dynamical nature of these acoustic modes that contributes to the Casimir effect, we believe the width of the temperature transition should be on the order of $\Delta T \approx \SI{10}{\angstrom^{1.5}}h^{-1.5}T_\mathrm{BKT}=\SI{0.45}{K}$, when scaled from the existing literature values~\cite{Ganshin06}. When comparing to the experiment data, we did not observe the critical feature measured in the existing literature about \SI{0.45}{K} around the transition temperature. Therefore, in the fitting, the functional form we choose to extract the far-from-transition behavior exhibits an abrupt change at the transition temperature and does not include the critical feature. Specifically, we took the fitting function, $f(T<T_0)=A(T/T_0)\mathrm{Re}[\varepsilon(l,T)^{-1}]; f(T>T_0)=0$, to capture both the overall amplitude $A$ of the Casimir effect, and the temperature scaling below the transition temperature $T_0=\SI{1.42}{K}$. The $\varepsilon$ dependence only provides a guide to the eye near the transition temperature. Note that the transition temperature deviates from the dissipation fit value by about 20\%, which could stem from a fitting error, or a change of film thickness after one thermal cycle of the fridge. 

As mentioned in the main text, the available theory fails to describe the rapid drop in resonance frequency that we observe above 1.42K. We have a few possible alternatives that could cause an abrupt effect. The rapid drop may be caused by an effect separate from the thickness dependence of the helium film. As one possibility, there might be some type of phase transition inside the vdW-striction interfaces in our device, perhaps associated with a layer of helium trapped there, which might cause a change in the steady-state tensile stress sufficient to account for the change in frequency.   Alternatively, there could be a transition in the external helium reservoir, which could lead to a relatively rapid change in chemical potential. 

Since we do not yet have sufficient understanding of the physics that might be involved in these alternate possibilities, we can not completely rule them out. However, we believe the Casimir physics is likely the cause of the frequency increase below 1.42K, due to the order-of-magnitude matching behavior with the observed data deep in the superfluid and normal-fluid phases. 

In the following, we describe the literature model that derives the Casimir expression scaling and show that, to first order, it is film thickness-independent when it's deep in the superfluid and normal-fluid phases.  We start from the most simplified case where all the acoustic modes are populated by thermal energy, and later discuss the impact of quantum fluctuations when the thickness $d$ (instead of $h$ in the previous section to be consistent with literature conventions) approaches a mono-layer level. 

\subsection*{Surface-Mode Casimir Energy}

In the superfluid phase, different from the normal fluid phase, undulations of the surface at \(z=d\) can propagate over long distances. We want a phase field \(\phi_{\mathrm{s}}\) such that (i) the normal velocity vanishes at the substrate, and (ii) matches the surface motion at \(z=d\). The minimal-energy solution satisfying \(\nabla^2\phi_{\mathrm{s}}=0\) and these boundary conditions is
\begin{equation*}
    \phi_{\mathrm{s}}(\mathbf{q}) =
    \frac{m}{\hbar}\,h_{\mathbf{q}}\,
    \frac{\cosh(q z)}{q\sinh(q d)}\,e^{i\mathbf{q}\cdot\mathbf{R}},
\end{equation*}
where \(q = |\mathbf{q}|\) and only lives on the surface plane (2D modes). This is the velocity potential associated with third-sound modes in the film, similar to the ones that we excite using oscillator motions.

These solutions allow us to calculate the kinetic energy
\(
E_{\mathrm{kin}} = \frac{\rho_s}{2}\int d^3x\,\mathbf{v}_s^2
\),
and the resulting surface-mode contribution to the free energy (only $d$-dependent terms)
\begin{equation}
\frac{F_{\mathrm{surf}}^{\mathrm{SF}}(d)}{A}
= \frac{k_B T}{2}\int\!\frac{d^2q}{(2\pi)^2}\,
\ln\!\bigl[\tanh(qd)\bigr].
\end{equation}
Evaluating the integral yields
\begin{equation*}
\frac{F_{\mathrm{surf}}^{\mathrm{SF}}(d)}{A}
= -\frac{7\,k_B T}{64\pi d^2}\,\zeta(3),
\end{equation*}
and the corresponding surface Casimir pressure is
\begin{equation}
P_{\mathrm{surf}}^{\mathrm{SF}}(d)
= -\frac{7\,k_B T}{32\pi d^3}\,\zeta(3)\approx -0.1\frac{k_B T}{d^3}.
\label{eq:PsurfSF}
\end{equation}

The calculations of Ref.~\cite{Zandi04} include a contribution to the Casimir force from bulk modes, which they find to be approximately 50\% of the surface contribution.  This gave a total Casimir pressure of
\begin{equation}
P_{\mathrm{tot}}^{\mathrm{SF}}(d)\approx\;- \Theta_0 \,\frac{k_B T}{d^3},\label{eq:Emp_Casimir}
\end{equation} 
with $\Theta_0 = 0.15$.  Experimental values for the thick films they considered were roughly $\Theta_0 \approx 0.3$.

For films whose thickness is of the order of one atomic layer, the modes contributing to the bulk Casimir effect no longer exist.  At the same time,   the surface contribution, which is dominated by fluctuations with a wavelength comparable to $d$  will presumably be modified significantly. Nevertheless, we may use Eq.~(\ref{eq:Emp_Casimir}), with $\Theta_0 = 0.3$, as an order-of-magnitude approximation to the superfluid thinning effect far from $T_\mathrm{BKT}$.

\subsection*{Quantum fluctuation corrections}

For a single harmonic oscillator of frequency $\omega$, the exact quantum free energy is
\begin{equation*}
F(\omega)
=
\frac{\hbar\omega}{2}
+
k_B T
\ln\!\bigl(1 - e^{-\beta\hbar\omega}\bigr),
\end{equation*}
Quantum corrections become important for frequencies such that $\hbar \omega > k_B T $. The dominant contributions to the surface Casimir effect come from fluctuations with wavelengths on the scale of the film thickness.   This gives a frequency $\omega*$ with 
\begin{equation}
T^* = \frac {\hbar \omega^* } {k_B}\approx0.33\,\mathrm{K}\left(\frac{c_3}{100\,\mathrm{m/s}}\right)\left(\frac{1\,\mathrm{nm}}{d}\right).
\end{equation}
Thus, quantum corrections should be at most marginally important for the temperatures of interest to us.

In our analysis, we directly use the thermal Casimir results in our fitting procedure, and use a geometric factor $\Theta_0$ to capture all the potential small correction factors from the differential properties of the reservoir surface and the contribution from unidentified acoustic modes, neglecting the offset from the quantum correction.

\subsection{\textsuperscript{4}He thickness analysis}

The universal jump predicts that the critical temperature is connected to the surface superfluid density: $n_s(T_\mathrm{BKT})=\frac{2m^2k_BT_\mathrm{BKT}}{\pi \hbar^2}$. We are able to backtrack from the $T_\mathrm{BKT}=\SI{1.2}{K}$ result to the superfluid thickness, which is about 0.9 atomic layer thick. Based on this thickness, we can estimate the third sound velocity $c_3=h_0/\rho_\mathrm{3D}\Pi'(h_0)$ in the vicinity of \SI{100}{m/s}, where $h_0$ is the liquid helium thickness, $\rho_\mathrm{3D}$ the density, and $\Pi'(h_0)=A_H/(2\pi h_0^4)$ the disjoining pressure. Given this acoustic velocity, our device dimensions, and resonator frequencies, we simulate and highlight the effective velocity participation ratio in Fig.\ref{fig:SI_SF}, which is then combined with the mass participation ratio (also thickness dependent) to calculate excess dissipation. We find that the dissipation is consistent within a factor of two. Fitting to the frequency shift data also requires a mass participation ratio, which leads to a literature-consistent geometric factor $\Theta\sim 0.3$. Given that the matching of both theoretical prediction to the experimental results is thickness sensitive, we believe we obtained a good estimation of the film thickness.

In terms of possible sources of such a thin He film in the fridge, we find a few plausible possibilities. But most importantly, we did not intentionally inject He gas into the fridge. We find that a total of \SI{6e18}{atoms} is consistent with the observed thickness. When these atoms condense and form liquid helium (density \SI{2.2e28}{atoms/m^3}), assuming 0.5-\SI{1}{m^2} of surface area at the mixing stage, we have an effective thickness of 0.3-\SI{0.6}{nm}, which is around 1-2 layers of He coverage. 

First, we want to estimate whether the gas condensation due to cryo-pumping alone can explain this amount of He atoms. The cool down starts at \SI{5e-3}{mbar}, which corresponds to about \SI{6e18}{atoms} in a \SI{50}{L} volume. In reality, most of the residue gases would likely not be He~\cite{KLOPFER19607}. Secondly, a similar amount of He atoms can be accumulated with a finite leakage rate around \SI{e-8}{mbar~L/s} with week-long operations. Such a source could be the metal valves for He circulation, or temporary thermal-expansion gaps during cooldowns, or just the external gas influx due to the seal leakage of the vacuum can. Since there are multiple layers of temperature shields, only He and hydrogen atoms can arrive at the sample stage. Further, as the He polarizability is concentrated near its 20-eV bandgap, where wide-gap diamond retains substantial dielectric response while metals behave almost transparent~\cite{Zaremba77}, there will be more He accumulation on the diamond surface compared to background surfaces~\cite{Vidali83,Vidali1983}. Lastly, the silver paste underneath the diamond is porous with surface area 1-\SI{10}{m^2/g}, potentially acting as a gather reservoir of He; diamond, in contrast, is about \SI{e-3}{m^2/g}. Since this thickness will only cause a thermal bridge to higher temperature stages with heating rate of about $\SI{0.1}{\micro\watt}$, considerably smaller than the typical cooling power at \SI{10}{mK}, we believe that this amount of He is not critical for fridge operation, so could be from various different sources that are hard to account for.

\bibliographystyle{apsrev4-2}
\bibliography{refs}

\end{document}